\documentclass[conference,a4paper]{IEEEtran}

\usepackage[utf8]{inputenc} 
\usepackage[T1]{fontenc}
\usepackage{url}
\usepackage{ifthen}
\usepackage{cite}
\usepackage[cmex10]{amsmath} 
\usepackage{amssymb}
\usepackage[linesnumbered,ruled]{algorithm2e}
\usepackage{graphicx}

\newtheorem{thm}{Theorem}[section]
\newtheorem{lem}[thm]{Lemma}

\newtheorem{defn}[thm]{Definition}

\newtheorem{exmp}[thm]{Example}

\newtheorem{condition}{Condition}

\begin{document}
\title{On Optimal Index Codes for Interlinked Cycle Structures with Outer Cycles}
\author{\IEEEauthorblockN{Shanuja Sasi and B. Sundar Rajan} 
\IEEEauthorblockA{Dept. of Electrical Communication Engg., Indian Institute of Science Bangalore, Karnataka, India - 560012\\
			Email: \{shanuja,bsrajan\}@iisc.ac.in}}	
\maketitle
 \thispagestyle{plain}
\pagestyle{plain}
	
\begin{abstract}

For index coding problems with special structure on the side-information graphs called Interlinked Cycle (IC) structures index codes have been proposed in the literature (C. Thapa, L. Ong, and S. Johnson, ``Interlinked Cycles for Index Coding: Generalizing Cycles and Cliques", in \textit{IEEE Trans. Inf. Theory, vol. 63, no. 6, Jun. 2017} with a correction in ``Interlinked Cycles for Index Coding: Generalizing Cycles and Cliques", in arxiv (arxiv:1603.00092v2 [cs.IT] 25 Feb 2018)). In this paper we consider a generalization of IC structures called {\it IC structures with interlocked outer cycles}.  For IC structures with interlocked outer cycles we show that the optimal length (also known as the minrank of the index coding problem) depends on the maximum number of disjoint outer cycles. We give two sufficient conditions such that if any of these is satisfied then we 
 provide explicit optimal index code construction. The conditions mentioned above are shown to be not necessary by an explicit example.
\end{abstract}
\section{INTRODUCTION}
\subsection{Background and Preliminaries}
In a single sender Index Coding problem with Side Information (ICSI), there will be a unique source having a set of $P$ messages $\mathcal{M} = \{x_{1},x_{2},...,x_{P}\}$ and $N$ receivers $\mathcal{R} =\{\mathcal{R}_{1},\mathcal{R}_{2},...,\mathcal{R}_{N} \}$. Each receiver $\mathcal{R}_i$ is characterized by $(\mathcal{W}_i,\mathcal{K}_i)$, where $\mathcal{W}_i \subseteq \mathcal{M}$ is the set of messages it demands and $\mathcal{K}_i \subseteq \mathcal{M}$ is the set of messages it knows a priori which is also the side information possessed by it, where $\mathcal{W}_i\cap\mathcal{K}_i=\emptyset$, i.e, it doesn't demand anything which it already has. The sender knows the side information available to the receivers. Each receiver should get the messages it requested from the sender with minimum number of transmissions. An ICSI is identified by $(\mathcal{M},\mathcal{R})$. An index code for the ICSI ($\mathcal{M},\mathcal{R}$) consists of an encoding function for the sender, $\mathcal{E} : \mathbb{F}_{q}^P \rightarrow \mathbb{F}_{q}^l$ and a set of decoding functions $\mathcal{D}_i : \mathbb{F}_{q}^l \times \mathbb{F}_{q}^{|\mathcal{K}_i|} \rightarrow \mathbb{F}_{q}^{|\mathcal{W}_i|}$ for each $i \in \{1,2,..,N\} $, such that,
			\begin{equation}
				\mathcal{D}_i(\mathcal{E}(\mathcal{\mathcal{M}}),\mathcal{K}_i) = \mathcal{W}_i ,\forall i \in \{1,2,..,N\},
			\end{equation}
			where $l$ is the length of the index code and $\mathbb{F}_q$ is a finite field with $q$ elements. The objective is to find the optimal index code which has the smallest $l$ possible, such that each receiver can decode the messages, which it demanded, from the codeword transmitted and the known messages. An index code is said to be linear if the encoding function $\mathcal{E}$ is linear. ICSI problem was introduced in \cite{BiK}.
		
			A detailed classification of ICSI is done by Ong and Ho \cite{OngHo} based on the demands and side information of the receivers. An ICSI problem is said to be \textit{unicast} if $\mathcal{W}_i \cap \mathcal{W}_j = \emptyset$ $\forall$ $i \neq j $ and it is \textit{single unicast} if $|\mathcal{W}_i|=1$. An ICSI problem is said to be \textit{uniprior} if $\mathcal{K}_i \cap \mathcal{K}_j = \emptyset$ $\forall$  $i \neq j $ and it is \textit{single uniprior} if $|\mathcal{K}_i|=1$. The most general setting of index coding instance is when there is no restriction on $\mathcal{W}_i$ and $\mathcal{K}_i$, which is said to be \textit{multicast multiprior}. Finding the optimal length for this case is known to be NP hard \cite{AHL}. 
			
			The notation $\lfloor v \rceil$ is used to represent the set $\{1,2,...,v\}$ for any integer $v$. We consider single unicast problem with $P=N$. Without loss of generality, let each receiver $R_i$ requests the message $x_i$. A \textit{Digraph} $G$ is specified by a set of vertices $V(G)$ and a set of arcs $E(G)$ connecting the vertices, denoted by $(V, E)$. An arc $(i, j) \in E(G)$, if the receiver $R_i$ has $x_j$ as side information. A \textit{path} is a digraph $P_{v_a,v_b}$, specified by $(V, E)$, where $V= \{v_a,v_1,v_2,...,v_t,v_b\}$ and $E=\{(v_a \rightarrow v_1),(v_1\rightarrow v_2),(v_2\rightarrow v_3),...,(v_t\rightarrow v_b)\}$. It is represented by $P_{v_a,v_b} = (v_a \rightarrow v_1 \rightarrow v_2 \rightarrow...\rightarrow v_t \rightarrow v_b)$. Two paths from $v_a$ to $v_b$ are said to be distinct, if the two paths do not have any vertices in common other than $v_a$ and $v_b$. 
			
			The \textit{optimal broadcast rate} for an index coding problem $G$ is defined as $$\beta(G) = \inf_{t} {\beta_t(G)},$$ where $\beta_t(G)$ is the minimum number of index code symbols to be transmitted, in order to satisfy the demand of all the receiver, per t-bit messages. A \textit{Maximum Induced Acyclic Sub-Digraph (MAIS)} is the induced acyclic sub-digraph formed by removing minimum number of vertices from the digraph $G$. $MAIS(G)$ is the order of the $MAIS$. For any digraph $G$, $MAIS(G)$ is the lower bound on the optimal broadcast rate, i.e, $\beta(G) \geq MAIS(G)$.
			
			In \cite{TOJ1,TOJ2} the authors consider a digraph $G$ with $N$ vertices and define an Interlinked Cycle (IC) structure as follows. Let $V$ be the set of $N$ vertices. Let there be a vertex set $V_I$, such that it has $K$ vertices, $V_I=\{1,2,...,K\}$, where for all ordered pair of vertices ($i$, $j$), there exists exactly one path from $i$ to $j$ which doesn't include any other vertices in $V_I$ except $i$ and $j$, where $i,j \in V_I$ and $i \neq j$. Such a vertex set $V_I$ is called inner vertex set and the vertices in $V_I$ is called inner vertices. All other vertices in $G$ other than inner vertices are called non-inner vertices and are included in the vertex set $V_{NI}$. A path in which only the first and the last vertices are from $V_I$, and they are distinct, is called an I-path. If the first and the last vertices are same, then it is called an I-cycle.   
			
			There exists a directed rooted tree from every vertex in $V_I$, denoted by $T_i, $ with any inner vertex $i$ as the root vertex and all other vertices in $V_I$ other than $i$ as leaves. If the digraph $G$ satisfies the following four conditions, then it is called an Interlinked Cycle structure, denoted by $G_K$. 
\begin{itemize}
\item \textit{Condition 1:} The union of all the $K$ rooted trees, each one rooted at unique inner vertex, should form the digraph $G$.
\item \textit{Condition 2:} There is no I-cycle in $G_K$.
\item \textit{Condition 3:} For all ordered pair of vertices ($i$, $j$), there exists exactly one path from $i$ to $j$ which doesn't include any other vertices in $V_I$ except $i$ and $j$, where $i,j \in V_I$ and $i \neq j$.
\item \textit{Condition 4:} No cycle containing only non-inner vertices.
\end{itemize}   
	In \cite{TOJ1,TOJ2}, the authors have proposed an index code construction scheme for an IC structure $G_K$ with $K$ inner vertices and a total of $N$ vertices. The code construction is as follows:
\begin{itemize}
\item A coded symbol is obtained by bitwise XOR of messages requested by all the inner vertices in $V_I$.
\item A coded symbol is obtained by the bitwise XOR of the message requested by $j$ and the messages requested by all its out-neighborhood vertices in $G_K$, for all $j \in V_{NI}$.
\end{itemize}
The code constructed by the above procedure has $N-K+1$ coded symbols.

The cycles containing only the non-inner vertices are called outer cycles. The IC structures satisfying only the first three conditions are called IC structures with outer cycles. This paper deals with IC structures with outer cycles. In \cite{VaR}, optimal length index code is provided for IC structures with one cycle among non-inner vertex set. In \cite{ViR}, an index code of length $N-K+1$ is provided for IC structures with arbitrary number of outer cycles with the issue of optimality of the index codes left open. In this paper we deal with  optimal index code constructions for IC structures with any number of outer cycles satisfying two conditions (given in Section \ref{sec2}) calling such IC structures as {\it IC structures with interlocked outer cycles}. 		


\subsection{Our Contributions}
The contributions in this paper may be highlighted as follows.
\begin{itemize}
\item  For IC structures with interlocked outer cycles we show that the optimal length depends on the maximum number of disjoint outer cycles. To be precise, if $t$ denotes the the maximum number of disjoint outer cycles then the optimal length is $N-K+2-t.$ 
\item We provide explicit optimal index code construction for the cases where either of the two conditions- Condition \ref{cond: ncyc1} and \ref{cond: ncyc2} given in Section \ref{main_results}  is satisfied. 
\item The conditions mentioned above are shown to be not necessary by an explicit example.
\item From the code construction in \cite{ViR}, the length of the index code obtained for any IC structure with cycles among non-inner vertex set is $N-K+1$. Algorithm \ref{algo2} in this paper gives an index code of length $N-K+2-t$, where $t$ is the maximum number of disjoint cycles in $\cal{C}$, for IC structures with interlocked outer cycles which obey either Condition \ref{cond: ncyc1} or Condition \ref{cond: ncyc2}. There is a difference of $t-1$ in the length of index codes. We will prove in Section \ref{main_results} that the index code obtained by Algorithm \ref{algo2} is optimal. Table $1$  illustrates the difference in the length of code obtained using Algorithm \ref{algo2} and the code constructed in \cite{ViR}.
 \end{itemize}
\begin{center}
                \begin{tabular}{ | p{2cm} | p{2.5cm} | p{2.5cm}  |}
                        \hline
                        \multicolumn{3}{|c|}{\textbf{Table 1}} \\ \hline
                        \multicolumn{3}{|p{8cm}|}{Table that illustrates the difference in the length of code obtained using our algorithm and previously known one.} \\ \hline
                        \textit{Examples}& \textit{Length of the code constructed by Algorithm \ref{algo2}}  &  \textit{Length of the code constructed in \cite{ViR} } \\ \hline
                        Example \ref{exmp: n_cat2} &12 & 12 \\ \hline
                        Example \ref{exmp: n_cat1} & 9 & 9 \\ \hline
                        Example \ref{exmp: n_eg3} &14 & 14  \\ \hline
                        Example \ref{exmp: n_eg4} & 12 & 13 \\  \hline
                        Example \ref{n_cyc_example1}  & 25 & 26 \\      \hline
                        Example \ref{exm: n_cyc_example2} & 22 & 22 \\  \hline
                        Example \ref{exm: n_cyc_example5} & 17 & 19 \\  \hline

                \end{tabular}

                        \label{tab1}
\end{center}

~ \\

\section{INTERLOCKED OUTER CYCLES}
\label{sec2}
\begin{defn}
\textit{Interlocked Outer Cycles:} Let there exist $n$ number of cycles, $C_j$ where $j \in \lfloor n \rceil$, among non-inner vertex set. The set of all cycles among non-inner vertices which obeys the two conditions below is called Interlocked Outer Cycles. Let $\cal{C}$ be the set of all cycles, $\cal{C}$ $=\{C_1,C_2,...,C_n\}$.
\begin{itemize}
\item \textit{Interlocking Condition (ILC):} If any cycle $C_i$ intersects any other cycle $C_k$, then $C_i$ and $C_k$ have either a single vertex or exactly a path in common, for any $i,k \in \lfloor n \rceil$. 
\item \textit{Central Cycle Condition (CCC):} A cycle $C_c \in \cal{C}$ such that all other cycles in $\cal{C}$ have atleast a single vertex in common with $C_c$ and the intersection of any two cycles in $\cal{C}$, other than $C_c$, has atleast a single vertex in common with $C_c$ is called a Central Cycle. There exists atleast one central cycle in $\cal{C}$.
\end{itemize}
\end{defn}

Let $V_{C_j}$ represents the subset of non-inner vertices which forms the cycle $C_j$, where $j \in \lfloor n \rceil$. Let $\hat{C}=\cup_{i=1}^{n} C_i, V_{OC} = \cup_{i=1}^{n} V_{C_i}$ and $V_{i,j}=V_{C_i} \cap V_{C_j}$, for any $i,j \in \lfloor n \rceil$. Let $\tilde{V}_{C_j}$ be the subset of vertices in $V_{C_j}$ which doesn't intersect any other cycles in $\cal{C}$, $\tilde{V}_{C_j}= V_{C_j} \backslash (\cup_{i \neq j,i=1}^{n} V_{i,j}$), where $j \in \lfloor n \rceil$.

	\begin{figure}[!t]
		\centering
		\includegraphics[width=15pc]{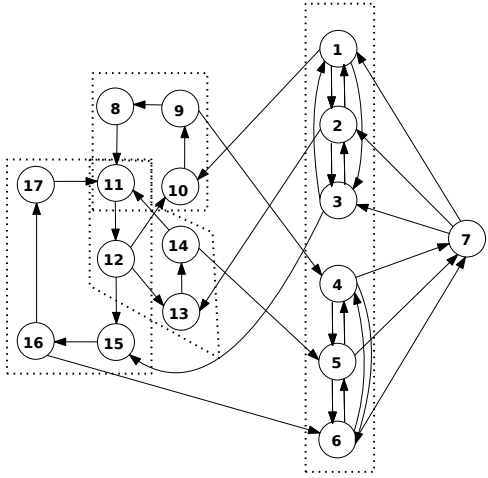}
		\caption{An IC structure $G_6$ with the path from the vertex $11$ to $12$ common to all the cycles in $C$.}
		\label{fig: n_cat2}
	\end{figure}
	
	\begin{exmp}
		Consider the IC structure $G_{6}$ shown in Fig. \ref{fig: n_cat2}. For this IC structure $N=17, K=6$, $V_I=\{1,2,3,4,5,6\}$, $V_{NI}=\{7,8,...,17\}$. Interlocked outer cycles are $C_1,C_2$ and $C_3$ with $V_{C_1}=
		\{11,12,15,16,17\},V_{C_2}=\{11,12,10,9,8\},V_{C_3}=\{11,12,13,14\} , V_{1,2}=\{11,12\}, V_{1,3}=\{11,12\}$ and $V_{2,3}=\{11,12\}$. All the three cycles have the path from the vertex $11$ to $12$ in common. All the three cycles are central cycles.
		\label{exmp: n_cat2}
\end{exmp}
	\begin{figure}[!t]
		\centering
		\includegraphics[width=15pc]{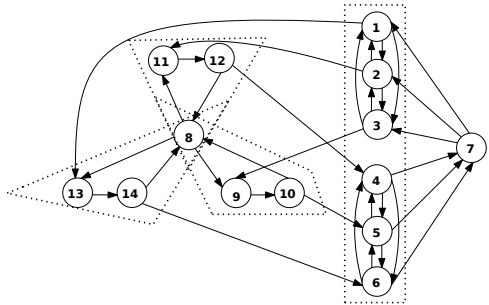}
		\caption{An IC structure $G_6$ with exactly a vertex $8$ common to all the cycles in $C$.}
		\label{fig: cat1}
	\end{figure}
	\begin{exmp}
		Consider the IC structure $G_{6}$ shown in Fig. \ref{fig: cat1}. For this IC structure $N=14, K=6$, $V_I=\{1,2,3,4,5,6\}$, $V_{NI}=\{7,8,...,14\}$. Interlocked outer cycles are $C_1,C_2$ and $C_3$ with $V_{C_1}=
		\{8,9,10\},V_{C_2}=\{8,11,12\}$ and $V_{C_3}=\{8,13,14\}$. All of them have the vertex $8$ in common. All the three cycles are central cycles. 
		\label{exmp: n_cat1}
	\end{exmp}
	
		\begin{figure}[!t]
			\centering
			\includegraphics[width=15pc]{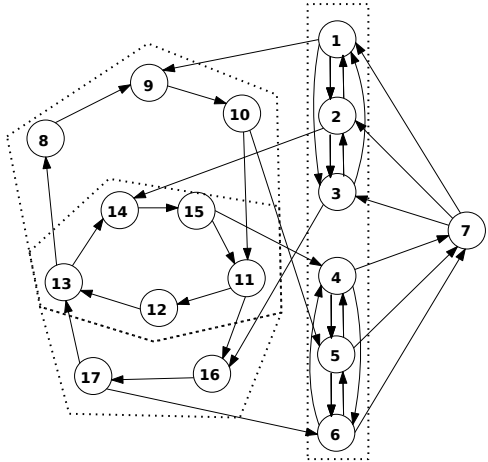}
			\caption{A digraph $G_6$ which is not an IC structure.}
			\label{cat_3}
		\end{figure}
		\begin{exmp}
			Consider the digraph $G_{6}$ shown in Fig. \ref{cat_3}. For this digraph $N=17, K=6$, $V_I=\{1,2,3,4,5,6\}$, $V_{NI}=\{7,8,...,17\}$. It has four cycles $C_1,C_2,C_3$ and $C_4$, containing only non-inner vertices with $V_{C_1}=
			\{8,9,10,11,12,13\},V_{C_2}=\{11,12,13,14,15\}, V_{C_3}=\{8,9,10,11,16,17,13\}$ and $V_{C_4}=\{14,15,11,16,17,13\}$. It has two I-paths from the vertex $1$ to $4$, $P_{1,4}^{1}=(1  \rightarrow 9 \rightarrow 10 \rightarrow 11 \rightarrow 12 \rightarrow 13 \rightarrow 14 \rightarrow 15 \rightarrow 4)$ and $P_{1,4}^{2}=(1 \rightarrow 9 \rightarrow 10 \rightarrow 11 \rightarrow 16 \rightarrow 17 \rightarrow 13 \rightarrow 14 \rightarrow 15 \rightarrow 4)$. Hence it is not an IC structure. 

The cycles $C_1$ and $C_4$ intersect but the intersection has neither a single vertex nor exactly a path in common. Hence it disobeys ILC. We conjecture that whenever ILC condition is violated, the digraph  will not be an IC structure.
			 \label{exmp: cat_3}
\end{exmp}

\begin{figure}[!t]
				\centering
				\includegraphics[width=15pc]{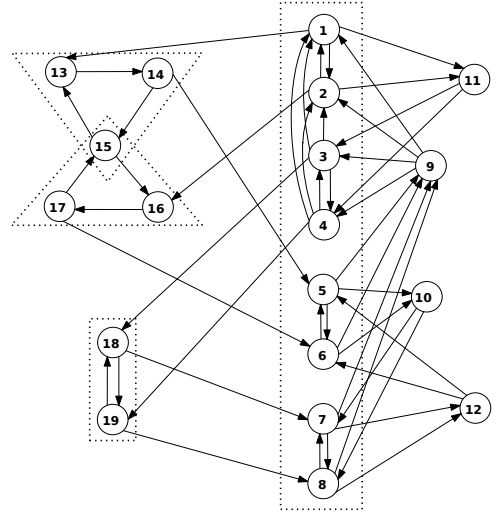}
				\caption{An IC structure without interlocked outer cycles.}
				\label{fig: n_cyc_CCC_violated}
\end{figure}
				
\begin{exmp}
Consider the IC structure $G_{8}$ shown in Fig. \ref{fig: n_cyc_CCC_violated}. For this IC structure $N=19, K=8$, $V_I=\{1,2,3,4,5,6,7,8\}$, $V_{NI}=\{9,10,...,19\}$. It has three cycles $C_1,C_2$ and $C_3$, containing only non-inner vertices with $V_{C_1}=
				\{13,14,15\},V_{C_2}=\{15,16,17\}$ and $V_{C_3}=\{18,19\}$. $C_1$ and $C_2$ intersect and the intersection has exactly a vertex $15$ in common. Hence it obeys ILC. None of the cycles in $C$ are central cycles. Hence it disobeys CCC. Therefore the set of all cycles $\{C_1,C_2,C_3\}$ is not interlocked outer cycles. \label{exmp: n_cyc_CCC_violated}
			\end{exmp}
	
		\begin{figure}[!t]
			\centering
			\includegraphics[width=20pc]{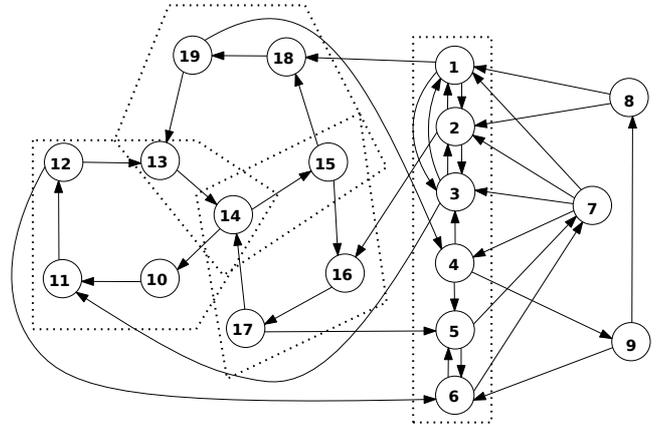}
			\caption{An IC structure $G_6$ with the vertex $14$ in common with all the cycles in $C$ and the vertex $13$ in common with $C_1$ and $C_2$ only.}
			\label{fig: n_eg3}
		\end{figure}
		

		\begin{exmp}
			Consider the IC structure $G_{6}$ shown in Fig. \ref{fig: n_eg3}. For this IC structure $N=19, K=6$, $V_I=\{1,2,3,4,5,6\}$, $V_{NI}=\{7,8,...,19\}$. Interlocked outer cycles are $C_1,C_2$ and $C_3$ with $V_{C_1}=
			\{10,11,12,13,14\},V_{C_2}=\{13,14,15,18,19\},V_{C_3}=\{14,15,16,17\} , V_{1,2}=\{13,14\}, V_{1,3}=\{14\}$ and $V_{2,3}=\{14,15\}$. All the three cycles have the vertex $14$ in common. All the three cycles are central cycles.
			\label{exmp: n_eg3}
		\end{exmp}
		
		\begin{figure}[!t]
			\centering
			\includegraphics[width=17pc]{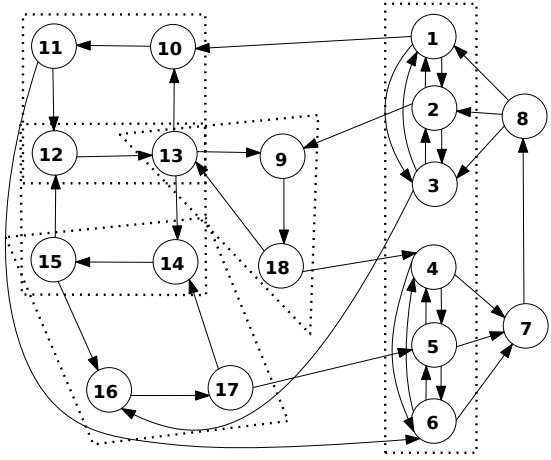}
			\caption{An IC structure $G_6$ which doesn't have any vertex in common with all the cycles in $C$.}
			\label{fig: n_eg4}
		\end{figure}
		
\begin{exmp}
			Consider the IC structure $G_{6}$ shown in Fig. \ref{fig: n_eg4}. For this IC structure $N=18, K=6$, $V_I=\{1,2,3,4,5,6\}$, $V_{NI}=\{7,8,...,18\}$. Interlocked outer cycles are $C_1,C_2,C_3$ and $C_4$ with $V_{C_1}=
			\{12,13,14,15\},V_{C_2}=\{10,11,12,13\},V_{C_3}=\{14,15,16,17\},V_{C_4}=\{13,9,18\}, V_{1,2}=\{13,12\}, V_{1,3}=\{14,15\},V_{1,4}=\{13\},V_{2,3}=\{13\}$ and $V_{2,4}=V_{3,4}= \phi$. All the four cycles do not have any vertex in common. The only central cycle is $C_1$.
			\label{exmp: n_eg4}
\end{exmp}
		
\begin{figure}[!t]
		\centering
		\includegraphics[width=18pc]{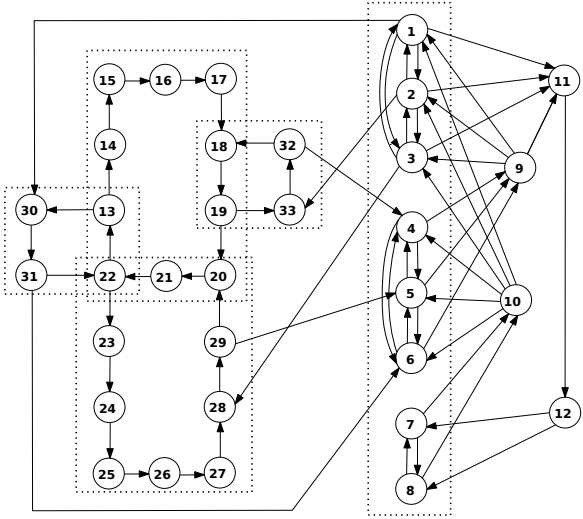}
		\caption{An IC structure $G_{8}$ which doesn't have any vertex in common with all the cycles in $C$.}
		\label{Fig: n_cyc_example1}
\end{figure}
\begin{exmp}
		Consider an IC structure $G_{8}$ shown in Fig. \ref{Fig: n_cyc_example1}. For this IC structure $N=33, K=8, V_I=\{1,2,3,4,5,6,7,8\}$ and $ V_{NI}=\{9,10,...,33\}$. Interlocked outer cycles are $C_1,C_2,C_3$ and $C_4$. $V_{C_1}=
		\{13,14,15,...,21,22\},V_{C_2}=\{20,21,22,...,28,29\}, V_{C_3}=\{13,30,31,22\},V_{C_4}=\{18,19,33,32\} ,V_{1,2}=\{20,21,22\}, V_{1,3}=\{22,13\},V_{1,4}=\{18,19\},V_{2,3}=\{22\}$ and $V_{2,4}=V_{3,4}= \phi$. There is no vertex in common with all the cycles. The only central cycle is $C_1$. \label{n_cyc_example1}	
\end{exmp}	

\subsection{Structure of IC Structures with Interlocked Outer Cycles}
In this subsection we describe two simple but important properties of IC structures with interlocked outer cycles.

\begin{lem}
		If all the $n$ cycles in $\cal{C}$ have atleast one vertex in common, then all the cycles in $\cal{C}$ are central cycles. \label{lem: multiple_PCC}
	\end{lem}
	
	\begin{IEEEproof}
		It is straight forward from the definition of a central cycle. If all the $n$ cycles in $\cal{C}$ have atleast one vertex $v_c$ in common, then if we take any cycle $C_i \in \cal{C}$, all other cycles have the vertex $v_c$ in common with $C_i$ and hence the intersection of any two cycles in $\cal{C}$ other than $C_i$ has $v_c$ in common with $C_i$. Hence all the cycles in $\cal{C}$ are central cycles, if all the $n$ cycles in $\cal{C}$ have atleast one vertex in common.
	\end{IEEEproof}
\begin{lem}
		For any central cycle $C_c \in \cal{C}$, there exists a path from any $i \in V_{C_k}$ to any $j \in V_{C_l}$, where $k,l \in \lfloor n \rceil$, and the path includes the vertices and arcs only from $C_k\cup C_l\cup C_c$. 
		\label{lem: existence_of_path_in_UoC} 
\end{lem}
	
\begin{IEEEproof} 
		For any $i,j \in V_{OC}$, the vertices $i$ and $j$ can be from the same cycle or different cycles. If $i$ and $j$ are from the same cycle, then there exists a path from $i$ to $j$ (part of that cycle) and the path includes the vertices and arcs from that cycle.
		
		If $i$ and $j$ are not in the same cycle, then let $i \in V_{C_k}, j\in V_{C_l}$, where $i,j \notin V_{k,l}$, $l\neq k$ and $k,l \in \lfloor n \rceil$. For this case, $V_{k,l}$ can be empty or non empty.
		
		\begin{itemize}	 	 	
			\item For $V_{k,l} \neq \phi$, let $P_{i,v_a} $ be the path from $i$ to some vertex $v_a \in V_{k,l}$ (part of the cycle $C_k$) and $P_{v_a,j}$ be the path from $v_a$ to $j$ (part of the cycle $C_l$). $(P_{i,v_a} \rightarrow P_{v_a,j})$ forms the path from $i$ to $j$ which includes the vertices and arcs from $C_k$ and $C_l$.	 	
			
			\item For $V_{k,l} = \phi$, since $C_c$ has either a single vertex or exactly a path in common with both $C_k$ and $C_l$, $V_{k,c}$ and $V_{l,c}$ will not be empty. Let $P_{i,v_a} $ be the path from $i$ to some vertex $v_a \in V_{k,c}$ (part of the cycle $C_k$), $P_{v_a,v_b}$ be the path from $v_a$ to some $v_b \in V_{l,c}$ (part of the cycle $C_c$) and $P_{v_b,j}$ be the path from $v_b$ to $j$ (part of the cycle $C_l$). $(P_{i,v_a} \rightarrow P_{v_a,v_b} \rightarrow P_{v_b,j})$ forms the path from $i$ to $j$ which includes the vertices and arcs from $C_k,C_c$ and $C_l$.		 	
		\end{itemize}
		
		Hence for any central cycle $C_c \in \cal{C}$, there exists a path from any $i \in V_{C_k}$ to any $j \in V_{C_l}$, where $k,l \in \lfloor n \rceil$, where the path includes the vertices and arcs from $C_k\cup C_l\cup C_c$. 	
\end{IEEEproof}

\section{Main Results}
\label{main_results}
	
	We consider  IC structures $G_K$ containing interlocked outer cycles with $N$ number of total vertices and $K$ number of inner vertices. 

        Let $V_I^{out}=\{k: k \in V_I$ and there exists a path from $k$ to $j$ for some $j \in V_{OC},$ where the path doesn't include any inner vertex other than $k\}$ and $V_I^{in}=\{k: k \in V_I$ and there exists a path from $j$ to $k$ for some $j \in V_{OC},$ where the path doesn't include any inner vertex other than $k\}$. Let $V_I^{*} = V_I \backslash (V_I^{in} \cup V_I^{out})$.

In this section, we prove that if an IC structure $G_K$, with interlocked outer cycles, satisfies either Condition \ref{cond: ncyc1} or Condition \ref{cond: ncyc2}  given below then the code constructed using Algorithm \ref{algo2} is optimal for $G_K$.


\begin{condition}
For any inner vertices $p,q \in V_I^{in} \cup V_I^{*}$ and $u,v \in V_I^{out}$, the I-path from $p$ to $q$ should not intersect the I-path from $u$ to $v$.\label{cond: ncyc1}
\end{condition}

\begin{condition}
For any inner vertices $p,q \in V_I^{out} \cup V_I^{*}$ and $u,v \in V_I^{in}$, the I-path from $p$ to $q$ should not intersect the I-path from $u$ to $v$.\label{cond: ncyc2}
\end{condition}

 	Consider Example \ref{exmp: n_cat2}. For this IC structure $V_I^{in} = \{4,5,6\}$, $V_I^{out} = \{1,2,3\}$ and $V_I^{*} = \phi$.
 	
 	Consider Example \ref{exmp: n_cat1}. For this IC structure $V_I^{in} = \{4,5,6\}$, $V_I^{out} = \{1,2,3\}$ and $V_I^{*} = \phi$.
 	
 	Consider Example \ref{exmp: n_eg3}. For this IC structure $V_I^{in} = \{4,5,6\}$, $V_I^{out} = \{1,2,3\}$ and $V_I^{*} = \phi$.
 	
 	Consider Example \ref{exmp: n_eg4}. For this IC structure $V_I^{in} = \{4,5,6\}$, $V_I^{out} = \{1,2,3\}$ and $V_I^{*} = \phi$.
 	
 	Consider Example \ref{Fig: n_cyc_example1}. For this IC structure $V_I^{in} = \{4,5,6\}$, $V_I^{out} = \{1,2,3\}$ and $V_I^{*} = \{7,8\}$.
 	
\begin{lem}
		If there is a path from $i \in V_I$ to $j$ for some $j \in V_{OC}$, then there doesn't exist any path from $k$ to the same inner vertex $i$ for any $k \in V_{OC}$. Similarly if there is a path from $k$ to $i \in V_I$ for some $k \in V_{OC}$, then there doesn't exist any path from $i$ to $j$ for any $j \in V_{OC}$. Hence $V_I^{in} \cap V_I^{out} = \phi$.\label{lem: no_bi_path}
\end{lem}
	
\begin{IEEEproof}
	Let there exist a path from $i \in V_I$ to $j$ for some $j \in V_{OC}$ and a path from $k$ to the same inner vertex $i$ for some $k \in V_{OC}$. Let $P_{i,j}$ be the path from $i$ to $j$ and $P_{k,i}$ be the path $k$ to $i$. From Lemma \ref{lem: existence_of_path_in_UoC}, there exists a path from $j$ to $k$ including only the vertices and arcs from $\hat{C}$. Hence there exists a path $P_{j,k}$ from $j$ to $k$ including only the vertices and arcs from $\hat{C}$. $P_{i,j}$, $P_{j,k}$ and $P_{k,i}$ form an I-cycle with only one inner vertex $i$, which contradicts the definition of an IC structure. Hence if there is a path from $i \in V_I$ to $j$ for some $j \in V_{OC}$, then there doesn't exist any path from $k$ to the same inner vertex $i$ for any $k \in V_{OC}$.
		
		Similarly we can prove that if there is a path from $k$ to $i \in V_I$ for some $k \in V_{OC}$, then there doesn't exist any path from $i$ to $j$ for any $j \in V_{OC}$.
		
		For each $i \in V_I^{out}$, there is a path from $i$ to $j$ for some $j \in V_{OC}$. So there doesn't exist any path from $k$ to $i$ for any $k \in V_{OC}$. Similarly for each $i \in V_I^{in}$, there is a path from $j$ to $i$ for some $j \in V_{OC}$. So there doesn't exist any path from $i$ to $k$ for any $k \in V_{OC}$. Therefore, $V_I^{in} \cap V_I^{out} = \phi$.
\end{IEEEproof}
	
\begin{lem}
	The I-path between any two vertices in $V_I^{out}$ does not intersect any vertex in $V_{OC}$. Similarly the I-path between any two vertices in $V_I^{in}$ also does not intersect any vertex in $V_{OC}$. \label{lem: V_inV_out_disj_V_{OC}}
\end{lem}
	
\begin{IEEEproof}
		By Lemma \ref{lem: no_bi_path}, $V_I^{in} \cap V_I^{out} = \phi$. Therefore the I-path between any two vertices in $V_I^{out}$ does not pass through any of the cycles in $\cal{C}$ and the I-path between any two vertices in $V_I^{in}$ also does not pass through any of the cycles in $\cal{C}$. 
\end{IEEEproof}
	
\begin{figure}[!t]
		\centering
		\includegraphics[width=13pc]{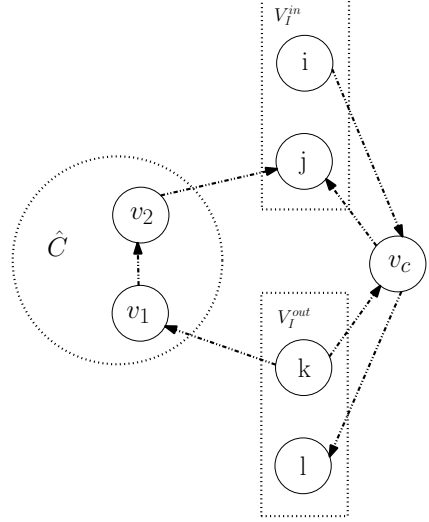}
		\caption{Illustration of the case explained in Lemma \ref{lem: disj_vout_vin}.}
		\label{n_in_out_disjoint}
\end{figure}
	
	
\begin{lem}
		The I-path between any two vertices in $V_I^{in}$ and the I-path between any two vertices in $V_I^{out}$ do not have any vertex $v_c \in V_{NI}\backslash V_{OC}$ in common. \label{lem: disj_vout_vin}
\end{lem}
	
\begin{IEEEproof} 
		Let $P_{i,j}$ represent the I-path from $i$ to $j$, where $i,j \in V_I^{in}$ and $P_{k,l}$ represent the I-path from $k$ to $l$, where $k,l \in V_I^{out}$. Let there exist a non-inner vertex $v_c \in V_{NI}\backslash V_{OC}$ common to both $P_{k,l}$ and $P_{i,j}$. Let $P_{k,v_1}$ be the path from $k$ to some vertex $v_1 \in V_{OC}$ and $P_{v_2,j}$ be the path from some vertex $v_2 \in V_{OC}$ to $j$. From Lemma \ref{lem: existence_of_path_in_UoC}, there exists a path from any vertex in $V_{OC}$ to any other vertex in $V_{OC}$ including only the vertices and arcs from $\hat{C}$. Hence there exists a path $P_{v_1,v_2}$ from the vertex $v_1$ to the vertex $v_2$ including only the vertices and arcs from $\hat{C}$. $P_{k,v_1}$, $P_{v_1,v_2}$ and $P_{v_2,j}$ form an I-path from $k$ to $j$. The path from $k$ to $v_c$ (part of $P_{k,l}$) and the path from $v_c$ to $j$ (part of $P_{i,j}$) form another I-path from $k$ to $j$. Hence there are two I-paths from $k$ to $j$ (shown in Fig. \ref{n_in_out_disjoint}). This contradicts the definition of an IC structure, of not having more than one I-path. Hence, the I-path between any two vertices in $V_I^{in}$ and the I-path between any two vertices in $V_I^{out}$ do not have any vertex $v_c \in V_{NI}\backslash V_{OC}$ in common. 
\end{IEEEproof}
	
		
\begin{lem}
	If an IC structure $G_K$, with interlocked outer cycles, obeys either Condition \ref{cond: ncyc1} or Condition \ref{cond: ncyc2}, then we can partition $V_I$ into two disjoint sets $V_I^1$ and $V_I^2$, where $V_I^1 \cup V_I^2=V_I$, such that the IC structure $G_K^{1}$, which is the induced sub-digraph of $G_K$ formed with the vertices $V_I^1$ and all the vertices in the I-path from each vertex in $V_I^1$ to every other vertex in $V_I^1$, and the IC structure $G_K^{2}$, which is the induced sub-digraph of $G_K$ formed with the vertices $V_I^2$ and all the vertices in the I-path from each vertex in $V_I^2$ to every other vertex in $V_I^2$, are disjoint, and $G_K^{1}$ and $G_K^{2}$ are disjoint from the interlocked outer cycles. \label{lem: 2_disj_ICs}
\end{lem}
	
\begin{IEEEproof}
		Consider an IC structure $G_K$ with interlocked outer cycles which obeys either Condition \ref{cond: ncyc1} or Condition \ref{cond: ncyc2}. 
		
		  By Lemma \ref{lem: disj_vout_vin}, I-path between any two vertices in $V_I^{in}$ and the I-path between any two vertices in $V_I^{out}$ do not have any vertex $v_c \in V_{NI}\backslash V_C$ in common. Hence the IC structure which is the induced sub-digraph of $G_K$ formed with the vertices $V_I^{in}$ and all the vertices in the I-path from each vertex in $V_I^{in}$ to every other vertex in $V_I^{in}$ and the IC structure which is the induced sub-digraph of $G_K$ formed with the vertices $V_I^{out}$ and all the vertices in the I-path from each vertex in $V_I^{out}$ to every other vertex in $V_I^{out}$ are disjoint.
		  
		  Let $V_I^1=V_I^{*} \cup V_I^{in}$ and $V_I^2=V_I^{out}$, if the IC structure $G_K$ obeys Condition \ref{cond: ncyc1}. Since I-path between any two vertices in $V_I^1$ doesn't intersect I-path between any two vertices in $V_I^2$ if the IC structure $G_K$ obeys Condition \ref{cond: ncyc1}, the IC structure which is the induced sub-digraph of $G_K$ formed with the vertices $V_I^1$ and all the vertices in the I-path from each vertex in $V_I^1$ to every other vertex in $V_I^1$ and the IC structure which is the induced sub-digraph of $G_K$ formed with the vertices $V_I^2$ and all the vertices in the I-path from each vertex in $V_I^2$ to every other vertex in $V_I^2$ are disjoint.
  
		  If the IC structure $G_K$ obeys Condition \ref{cond: ncyc2}, then let $V_I^2=V_I^{*} \cup V_I^{out}$ and $V_I^1=V_I^{in}$. Since I-path between any two vertices in $V_I^1$ doesn't intersect I-path between any two vertices in $V_I^2$ if the IC structure $G_K$ obeys Condition \ref{cond: ncyc2}, the IC structure which is the induced sub-digraph of $G_K$ formed with the vertices $V_I^1$ and all the vertices in the I-path from each vertex in $V_I^1$ to every other vertex in $V_I^1$ and the IC structure which is the induced sub-digraph of $G_K$ formed with the vertices $V_I^2$ and all the vertices in the I-path from each vertex in $V_I^2$ to every other vertex in $V_I^2$ are disjoint.
		  
		  If the IC structure $G_K$ obeys both Condition \ref{cond: ncyc1} and Condition \ref{cond: ncyc2}, then let $V_I^1=V_I^{*} \cup V_I^{in}$, $V_I^2=V_I^{out}$ or $V_I^2=V_I^{*} \cup V_I^{out}$, $V_I^1=V_I^{in}$. Since I-path between any two vertices in $V_I^1$ doesn't intersect I-path between any two vertices in $V_I^2$, the IC structure which is the induced sub-digraph of $G_K$ formed with the vertices $V_I^1$ and all the vertices in the I-path from each vertex in $V_I^1$ to every other vertex in $V_I^1$ and the IC structure which is the induced sub-digraph of $G_K$ formed with the vertices $V_I^2$ and all the vertices in the I-path from each vertex in $V_I^2$ to every other vertex in $V_I^2$ are disjoint. 
		  
		  Let $G_K^{1}$ be the IC structure which is the induced sub-digraph of $G_K$ formed with the vertices $V_I^1$ and all the vertices in the I-path from each vertex in $V_I^1$ to every other vertex in $V_I^1$ and $G_K^{2}$ be the IC structure which is the induced sub-digraph of $G_K$ formed with the vertices $V_I^2$ and all the vertices in the I-path from each vertex in $V_I^2$ to every other vertex in $V_I^2$.
		  
		  By Lemma \ref{lem: V_inV_out_disj_V_{OC}}, I-path between any two vertices in $V_I^{out}$ does not intersect any vertex in $V_C$. Similarly I-path between any two vertices in $V_I^{in}$ also does not intersect any vertex in $V_C$. Hence I-path between any two vertices in $V_I^2$ does not intersect any vertex in $V_C$. Similarly I-path between any two vertices in $V_I^1$ also does not intersect any vertex in $V_C$. Therefore IC structures $G_K^{1}$ and $G_K^{2}$ are disjoint from the cycles in $C$. There is no other cycle among non-inner vertices, other than those in $C$, for the IC structure $G_K$. Hence IC structures $G_K^{1}$ and $G_K^{2}$ do not have any cycles among non-inner vertices.
 \end{IEEEproof}
		

  \subsection{Algorithm}
  
In this subsection we give an algorithm to construct an index code for IC structures with interlocked outer cycles obeying either Condition \ref{cond: ncyc1} or Condition \ref{cond: ncyc2}.

\begin{algorithm}
  	\caption{Algorithm to obtain an index code for IC structures with interlocked outer cycles which obey either Condition \ref{cond: ncyc1} or Condition \ref{cond: ncyc2}.}
  	
  	\SetKwInOut{Input}{Input}
  	
  	\Input{An IC structure $G_K$ with interlocked outer cycles.}
  	
  	\begin{enumerate}
  		 
        \item	Find $C^S$ which is a set of maximum number of disjoint cycles in $\cal{C}$. $|C^S|=t$, where $t$ is the maximum number of disjoint cycles in $\cal{C}$. For each $C_i \in C^{S}$, let $V_{C_i} =\{x_{i_1},x_{i_2},x_{i_3},...,x_{i_{N_{C_i}}}\}$ such that, $(x_{i_1} \rightarrow x_{i_2} \rightarrow x_{i_3} \rightarrow ... \rightarrow x_{i_{N_{C_i}}} \rightarrow x_{i_1} )$ forms the cycle $C_i$. 
  		
  		\item Partition $V_I$ into two disjoint sets $V_I^{1}$ and $V_I^{2}$, where $V_I^{1} \cup V_I^{2}=V_I$, such that the IC structure $G_K^{1}$, which is the induced sub-digraph of $G_K$ formed with the vertices $V_I^{1}$ and all the vertices in the I-path from each vertex in $V_I^{1}$ to every other vertex in $V_I^{1}$, and the IC structure $G_K^{2}$, which is the induced sub-digraph of $G_K$ formed with the vertices $V_I^{2}$ and all the vertices in the I-path from each vertex in $V_I^{2}$ to every other vertex in $V_I^{2}$, are disjoint, and $G_K^{1}$ and $G_K^{2}$ are disjoint from the interlocked outer cycles.
  		\item Let $V^{1}$ represents the set of all vertices in the IC structure $G_{K}^{1}$ and $V^{2}$ represents the set of all vertices in the IC structure $G_{K}^{2}$. Let $V_{NI}^{1}= V^{1} \backslash V_I^{1}$ and $V_{NI}^{2}= V^{2} \backslash V_I^{2}$.
  		
  		\item Use IC structure code construction given in \cite{TOJ1,TOJ2} for $G_K^1, G_K^2$ and all the cycles in $C^S$, i.e, 
  		\begin{enumerate} 	
  	  		\item An index code is obtained by the bitwise XOR of messages requested by all the vertices in the set $V_I^{1}$, i.e, $$ w_I^1 = \oplus_{i\in V_I^{1}} x_i.$$
  		
  		\item An index code is obtained by the bitwise XOR of messages requested by the vertices in the set $V_I^{2}$, i.e, $$ w_I^2 = \oplus_{i\in V_I^{2}} x_i.$$
  		
  		\item For each $C_i \in C^{S}$, an index code $\{ x_{i_1} \oplus x_{i_2}, x_{i_2} \oplus x_{i_3}, x_{i_3} \oplus x_{i_4},..., x_{i_{N_{C_i}-1}} \oplus x_{i_{N_{C_i}}}\}$ of length $N_{C_i}-1$ is obtained.
  		
  		\item For each $j \in V_{NI}^{1} $, an index code is obtained by the bitwise XOR of message requested by $j$ with the messages requested by its out-neighborhood vertices in the IC structure $G_K^1$ and for each $j \in  V_{NI}^{2}$, an index code is obtained by the bitwise XOR of message requested by $j$ with the messages requested by its out-neighborhood vertices in the IC structure $G_K^2$, i.e,\\ $w_j = x_j \oplus_{k \in N_{G_K^1}^{+}(j)}x_k, $ if $j \in V_{NI}^{1} $\\ $w_j = x_j \oplus_{k \in N_{G_K^2}^{+}(j)}x_k,$ if $j \in V_{NI}^{2}$\\ where $N_{G}^{+}(p)$ represents the set of out-neighborhood vertices of the vertex $p$ in the IC structure $G$.

  		\end{enumerate}
  		\item For the remaining vertices, i.e, for each $j \in V \backslash ((\cup_{C_k \in C^{S}}V_{C_k}) \cup V^{1} \cup V^{2}),$ the messages requested by those vertices are sent uncoded. $$w_j=x_j.$$
  		
  	\end{enumerate} \label{algo2}
\end{algorithm}
  \textit{Proof of Correctness of Algorithm \ref{algo2}:}
    Consider an IC structure $G_K$, with interlocked outer cycles, which obeys either Condition \ref{cond: ncyc1} or Condition \ref{cond: ncyc2}. By Lemma \ref{lem: 2_disj_ICs}, we can partition $V_I$ into two disjoint sets $V_I^{1}$ and $V_I^{2}$, where $V_I^{1} \cup V_I^{2}=V_I$, such that the IC structure $G_K^{1}$, which is the induced sub-digraph of $G_K$ formed with the vertices $V_I^{1}$ and all the vertices in the I-path from each vertex in $V_I^{1}$ to every other vertex in $V_I^{1}$, and the IC structure $G_K^{2}$, which is the induced sub-digraph of $G_K$ formed with the vertices $V_I^{2}$ and all the vertices in the I-path from each vertex in $V_I^{2}$ to every other vertex in $V_I^{2}$, are disjoint and $G_K^{1}$ and $G_K^{2}$ are disjoint from the interlocked outer cycles. 
    
    There is no other cycle among non-inner vertices, other than those in $C$, for the IC structure $G_K$. Hence IC structures $G_K^{1}$ and $G_K^{2}$ do not have any cycles among non-inner vertices. Therefore IC structure code construction given in \cite{TOJ1,TOJ2} is valid and decodable for $G_K^1, G_K^2$ and all the cycles in $C^S$.
    
    
  \begin{exmp}
	  Consider Example \ref{exmp: n_cat2}. All the three cycles have the vertices $11$ and $12$ in common. A set of maximum number of disjoint cycles in $\cal{C}$ is $C^S=\{C_2\}$. Let $V_I^{1}=\{1,2,3\}$ and $V_I^{2}=\{4,5,6\}$. The IC structure $G_6^{1}$, which is the induced sub-digraph of $G_6$ formed with the vertices $V_I^{1}$ and all the vertices in the I-path from each vertex in $V_I^{1}$ to every other vertex in $V_I^{1}$, and the IC structure $G_6^{2}$, which is the induced sub-digraph of $G_6$ formed with the vertices $V_I^{2}$ and all the vertices in the I-path from each vertex in $V_I^{2}$ to every other vertex in $V_I^{2}$, are disjoint and $G_6^{1}$ and $G_6^{2}$ are disjoint from the interlocked outer cycles.\\  
	   $V_{NI}^{1}=V_{NI}^{2}= \phi,\\ V^{1}=\{1,2,3\},\\ V^{2}=\{4,5,6\}$ and \\$V \backslash ((\cup_{C_k \in C^{S}}V_{C_k}) \cup V^{1} \cup V^{2})=\{15,16,17,13,14,7\}$.\\ Using Algorithm \ref{algo2}, 
		  \begin{enumerate}
		  	\item  $w_I^1 = x_1 \oplus x_2 \oplus x_3.$
		  	
		  	\item $ w_I^2 = x_4 \oplus x_5 \oplus x_6.$
		  	
		  	\item For $C_2 \in C^{S}$, an index code $\{ x_{10} \oplus x_9,x_9 \oplus x_8, x_8 \oplus x_{11}, x_{11} \oplus x_{12} \}$ of length $4$ is obtained.

		  	\item For each $j \in \{15,16,17,13,14,7\},$ $w_j=x_j.$
		  
		  \end{enumerate}
	\label{exm: 1_code}
 \end{exmp}

	
	\begin{exmp}
		Consider Example \ref{exmp: n_cat1}. All the three cycles have the vertex $8$ in common. A set of maximum number of disjoint cycles in $\cal{C}$ is $C^S=\{C_2\}$. Let $V_I^{1}=\{1,2,3\}$ and $V_I^{2}=\{4,5,6\}$. The IC structure $G_6^{1}$, which is the induced sub-digraph of $G_6$ formed with the vertices $V_I^{1}$ and all the vertices in the I-path from each vertex in $V_I^{1}$ to every other vertex in $V_I^{1}$, and the IC structure $G_6^{2}$, which is the induced sub-digraph of $G_6$ formed with the vertices $V_I^{2}$ and all the vertices in the I-path from each vertex in $V_I^{2}$ to every other vertex in $V_I^{2}$, are disjoint and $G_6^{1}$ and $G_6^{2}$ are disjoint from the interlocked outer cycles. \\
		$V_{NI}^{1}=V_{NI}^{2}= \phi,\\ V^{1}=\{1,2,3\},\\ V^{2}=\{4,5,6\}$ and \\$V \backslash ((\cup_{C_k \in C^{S}}V_{C_k}) \cup V^{1} \cup V^{2})=\{13,14,9,10,7\}$.\\ Using Algorithm \ref{algo2}, 
			\begin{enumerate}
				\item  $w_I^1 = x_1 \oplus x_2 \oplus x_3.$
				
				\item $ w_I^2 = x_4 \oplus x_5 \oplus x_6.$
				
				\item For $C_2 \in C^{S}$, an index code $\{ x_{8} \oplus x_{11},x_{11} \oplus x_{12} \}$ of length $2$ is obtained.

				\item For each $j \in \{13,14,9,10,7\},$ $w_j=x_j.$
				
			\end{enumerate}
	\label{exm: 2_code}
	\end{exmp}


	\begin{exmp}
		Consider Example \ref{exmp: n_eg3}. All the three cycles have the vertex $14$ in common. A set of maximum number of disjoint cycles in $\cal{C}$ is $C^S=\{C_2\}$. Let $V_I^{1}=\{1,2,3\}$ and $V_I^{2}=\{4,5,6\}$. The IC structure $G_6^{1}$, which is the induced sub-digraph of $G_6$ formed with the vertices $V_I^{1}$ and all the vertices in the I-path from each vertex in $V_I^{1}$ to every other vertex in $V_I^{1}$, and the IC structure $G_6^{2}$, which is the induced sub-digraph of $G_6$ formed with the vertices $V_I^{2}$ and all the vertices in the I-path from each vertex in $V_I^{2}$ to every other vertex in $V_I^{2}$, are disjoint and $G_6^{1}$ and $G_6^{2}$ are disjoint from the interlocked outer cycles. \\
		$V_{NI}^{1}= \phi,\\V_{NI}^{2}= \{7,9\},\\ V^{1}=\{1,2,3\},\\ V^{2}=\{4,5,6,7,9\}$ and\\ $V \backslash ((\cup_{C_k \in C^{S}}V_{C_k}) \cup V^{1} \cup V^{2})=\{10,11,12,16,17,8\}$. \\Using Algorithm \ref{algo2}, 
			\begin{enumerate}
				\item  $w_I^1 = x_1 \oplus x_2 \oplus x_3.$
				
				\item $ w_I^2 = x_4 \oplus x_5 \oplus x_6.$
				
				\item $V_{NI}^{1}\cup V_{NI}^{2}= \{7,9\}$.\\ $N_{G_K^{2}}^{+}(7)=\{4\}$ and \\$N_{G_K^{2}}^{+}(9)=\{6\}$.\\ $w_7=x_7 \oplus x_4$ and \\$w_9=x_9 \oplus x_6.$
				
				\item For $C_2 \in C^{S}$, an index code $\{ x_{13} \oplus x_{14}, x_{14} \oplus x_{15}, x_{15} \oplus x_{18}, x_{18} \oplus x_{19} \}$ of length $4$ is obtained.

				\item For each $j \in \{10,11,12,16,17,8\},$ $w_j=x_j.$
			\end{enumerate}
	\label{exm: 3_code}
	\end{exmp}


	\begin{exmp}
		Consider Example \ref{exmp: n_eg4}. All the four cycles do not have any vertex in common. A set of maximum number of disjoint cycles in $\cal{C}$ is $C^S=\{C_2,C_3\}$. Let $V_I^{1}=\{1,2,3\}$ and $V_I^{2}=\{4,5,6\}$. The IC structure $G_6^{1}$, which is the induced sub-digraph of $G_6$ formed with the vertices $V_I^{1}$ and all the vertices in the I-path from each vertex in $V_I^{1}$ to every other vertex in $V_I^{1}$, and the IC structure $G_6^{2}$, which is the induced sub-digraph of $G_6$ formed with the vertices $V_I^{2}$ and all the vertices in the I-path from each vertex in $V_I^{2}$ to every other vertex in $V_I^{2}$, are disjoint and $G_6^{1}$ and $G_6^{2}$ are disjoint from the interlocked outer cycles.
		\\ $V_{NI}^{1}= \phi,\\V_{NI}^{2}= \phi,\\ V^{1}=\{1,2,3\}, \\V^{2}=\{4,5,6\}$ and \\$V \backslash ((\cup_{C_k \in C^{S}}V_{C_k}) \cup V^{1} \cup V^{2})=\{9,18,7,8\}$.\\ Using Algorithm \ref{algo2}, 
			\begin{enumerate}
				\item  $w_I^1 = x_1 \oplus x_2 \oplus x_3.$
				
				\item $ w_I^2 = x_4 \oplus x_5 \oplus x_6.$

				\item For $C_2 \in C^{S}$, an index code $\{ x_{13} \oplus x_{10}, x_{10} \oplus x_{11}, x_{11} \oplus x_{12}\}$ of length $3$ is obtained.
				
				\item For $C_3 \in C^{S}$, an index code $\{ x_{14} \oplus x_{15},x_{15} \oplus x_{16},x_{16} \oplus x_{17}\}$ of length $3$ is obtained.

				\item For each $j \in \{9,18,7,8\},$ $w_j=x_j.$
			\end{enumerate}
	
	\label{exm: 4_code}
	\end{exmp}

	\begin{exmp}
		Consider Example \ref{n_cyc_example1}. All the four cycles do not have any vertex in common. A set of maximum number of disjoint cycles in $\cal{C}$ is $C^S=\{C_3,C_4\}$. Let $V_I^{1}=\{1,2,3\}$ and $V_I^{2}=\{4,5,6,7,8\}$. The IC structure $G_8^{1}$, which is the induced sub-digraph of $G_8$ formed with the vertices $V_I^{1}$ and all the vertices in the I-path from each vertex in $V_I^{1}$ to every other vertex in $V_I^{1}$, and the IC structure $G_8^{2}$, which is the induced sub-digraph of $G_8$ formed with the vertices $V_I^{2}$ and all the vertices in the I-path from each vertex in $V_I^{2}$ to every other vertex in $V_I^{2}$, are disjoint and $G_8^{1}$ and $G_8^{2}$ are disjoint from the interlocked outer cycles.
		\\ $V_{NI}^{1}= \phi,\\V_{NI}^{2}= \{9,11,12,10\},\\ V^{1}=\{1,2,3\},\\ V^{2}=\{4,5,6,7,8,9,10,11,12\}$ and\\ $V \backslash ((\cup_{C_k \in C^{S}}V_{C_k}) \cup V^{1} \cup V^{2})=\{14,15,16,17,20,21,23,24,25,26,27,28,29\}$.\\ Using Algorithm \ref{algo2}, 
			\begin{enumerate}
				\item  $w_I^1 = x_1 \oplus x_2 \oplus x_3.$
				
				\item $ w_I^2 = x_4 \oplus x_5 \oplus x_6 \oplus x_7 \oplus x_8.$
				
				\item $V_{NI}^{1} \cup V_{NI}^{2}= \{9,10,11,12\}$.\\ $N_{G_K^{2}}^{+}(9)=\{11\},\\ N_{G_K^{2}}^{+}(11)=\{12\},\\N_{G_K^{2}}^{+}(12)=\{7,8\}$ and\\ $N_{G_K^{2}}^{+}(10)=\{4,5,6\}$.\\ $w_9=x_9 \oplus x_{11},\\ w_{11}=x_{11} \oplus x_{12},\\ w_{12}=x_{12} \oplus x_{7} \oplus x_8$ and \\$w_{10}=x_{10} \oplus x_4 \oplus x_5 \oplus x_6.$
				
				\item For $C_3 \in C^{S}$, an index code $\{ x_{13} \oplus x_{30},x_{30} \oplus x_{31}, x_{31} \oplus x_{22} \}$ of length $3$ is obtained.
				
				\item For $C_4 \in C^{S}$, an index code $\{x_{18} \oplus x_{19},x_{19} \oplus x_{33},x_{33} \oplus x_{32}\}$ of length $3$ is obtained.

				\item For each $j \in \{14,15,16,17,20,21,23,24,25,26,27,28,29\},$ $w_j=x_j.$
			\end{enumerate}
	\label{exm: 5_code}
	\end{exmp}
 \subsection{Optimality} 
In this section we show that the code construction given by the algorithm in the previous subsection gives an optimal length code.
	\begin{thm}
		A lower bound on the optimal broadcast rate $\beta(G_K)$ for the IC structure $G_K$, with interlocked outer cycles is $N-K+2-t$, where $t$ is the maximum number of disjoint cycles in $\cal{C}$.
		\label{thm: lower_bound_obr}
	\end{thm}
	
	\begin{IEEEproof}
		Consider the interlocked outer cycles in the IC structure $G_K$. Find $C^S$ which is a set of maximum number of disjoint cycles in $\cal{C}$. $|C^S|=t$, where $t$ is the maximum number of disjoint cycles in $\cal{C}$. The minimum number of vertices to be removed from $V_{OC}$ to make the induced sub-digraph formed with the vertices in $V_{OC}$ acyclic is $t$. Let $S$ be a set of minimum number of vertices to be removed from $V_{OC}$ to make the induced sub-digraph formed with the vertices in $V_{OC}$ acyclic.
		
		Take any I-path which passes through some vertex $v_{k} \in S$. Let it be the I-path from $i$ to $j$, where $i \in V_I^{out}$ and $j\in V_I^{in}$. By removing the $t$ vertices in $S$, the induced sub-digraph containing all the vertices in $V_{C}$ and the induced sub-digraph containing all the vertices in the path from $i$ to $j$ and $j$ to $i$ have become acyclic. By removing the inner vertices $V_I \backslash \{i,j\}$, the induced sub-digraph containing all the vertices in $V \backslash (V_C \cup \{i,j\})$ has become acyclic. Hence the induced sub-digraph with $N-K+2-t$ vertices, $V_{NI} \cup \{i,j\} \backslash S$, is acyclic. Hence, $MAIS(G_K) \geq N-K+2-t$. The optimal broadcast rate $\beta(G_K)$ is lower bounded by $MAIS(G_K)$. Therefore, $\beta(G_K) \geq N-K+2-t$.
	\end{IEEEproof}
		
	\begin{thm}
		 A sufficient but not necessary condition for the code construction given in Algorithm \ref{algo2} to be optimal for an IC structure $G_K$, with interlocked outer cycles, is either Condition \ref{cond: ncyc1} or Condition \ref{cond: ncyc2}.
	\end{thm}
	
	\begin{IEEEproof}
		Consider an IC structure $G_K$ with interlocked outer cycles which obeys either Condition \ref{cond: ncyc1} or Condition \ref{cond: ncyc2}.

	  	   By Lemma \ref{lem: 2_disj_ICs}, we can partition $V_I$ into two disjoint sets $V_I^1$ and $V_I^2$, where $V_I^1 \cup V_I^2=V_I$, such that the IC structure $G_K^{1}$, which is the induced sub-digraph of $G_K$ formed with the vertices $V_I^{1}$ and all the vertices in the I-path from each vertex in $V_I^{1}$ to every other vertex in $V_I^{1}$, and the IC structure $G_K^{2}$, which is the induced sub-digraph of $G_K$ formed with the vertices $V_I^{2}$ and all the vertices in the I-path from each vertex in $V_I^{2}$ to every other vertex in $V_I^{2}$, are disjoint and $G_K^{1}$ and $G_K^{2}$ are disjoint from the interlocked outer cycles. The IC structures $G_K^{1},G_K^{2}$ and the cycles in $C^{S}$ are disjoint from one another.
	   
		   Use Algorithm \ref{algo2} to obtain an index code for the IC structure $G_K$. Let the total number of vertices in $G_K^{1}$ and $G_K^{2}$ be $N_1$ and $N_2$ respectively and number of inner vertices be $K_1$ and $K_2$ respectively, where $K_1+K_2=K$. Let the number of vertices in $C_k$ be $N_{C_k}$, for each $C_k\in C^{S}$. Then there are $N^{'}=N-N_1-N_2-\sum_{k:C_k \in C^{S}}N_{C_k}$ vertices remaining. Number of coded symbols transmitted is $(N_1 -K_1+1) + (N_2-K_2+1) + \sum_{k:C_k \in C^{S}}(N_{C_k}-1) + N^{'} =N-K+2-t $.
		
			Thus we have constructed an index code which satisfies the lower bound on the optimal broadcast rate (given in Theorem \ref{thm: lower_bound_obr}). Hence it is optimal.

			It is shown in Example \ref{exmp: VC1VC2_optimal} that even if an IC structure obeys neither Condition \ref{cond: ncyc1} nor \ref{cond: ncyc2}, code construction given by Algorithm \ref{algo2} can be optimal.
	\end{IEEEproof}
	
	    Consider Example \ref{exm: 1_code}. This IC structure obeys both Condition \ref{cond: ncyc1} and \ref{cond: ncyc2}.  The induced sub-digraph with vertices $V_{NI} \cup \{1,5\} \backslash \{11\}$ is acyclic. Therefore $MAIS(G_K) \geq 12$ and hence $\beta(G_K) \geq 12$. Number of transmitted symbols is $12$, which achieves the lower bound on the optimal broadcast rate. Hence this code is optimal.
	    
	    Consider Example \ref{exm: 2_code}. This IC structure obeys both Condition \ref{cond: ncyc1} and \ref{cond: ncyc2}.  The induced sub-digraph with vertices $V_{NI} \cup \{1,5\} \backslash \{8\}$ is acyclic. Therefore $MAIS(G_K) \geq 9$ and hence $\beta(G_K) \geq 9$. Number of transmitted symbols is $9$, which achieves the lower bound on the optimal broadcast rate. Hence this code is optimal.
	    
	    Consider Example \ref{exm: 3_code}. This IC structure obeys both Condition \ref{cond: ncyc1} and \ref{cond: ncyc2}.  The induced sub-digraph with vertices $V_{NI} \cup \{1,5\} \backslash \{14\}$ is acyclic. Therefore $MAIS(G_K) \geq 14$ and hence $\beta(G_K) \geq 14$. Number of transmitted symbols is $14$, which achieves the lower bound on the optimal broadcast rate. Hence this code is optimal.
	    
	    Consider Example \ref{exm: 4_code}. This IC structure obeys both Condition \ref{cond: ncyc1} and \ref{cond: ncyc2}.  The induced sub-digraph with vertices $V_{NI} \cup \{1,5\} \backslash \{13,14\}$ is acyclic. Therefore $MAIS(G_K) \geq 12$ and hence $\beta(G_K) \geq 12$. Number of transmitted symbols is $12$, which achieves the lower bound on the optimal broadcast rate. Hence this code is optimal.
	
		\begin{figure}[!t]
			\centering
			\includegraphics[width=17pc]{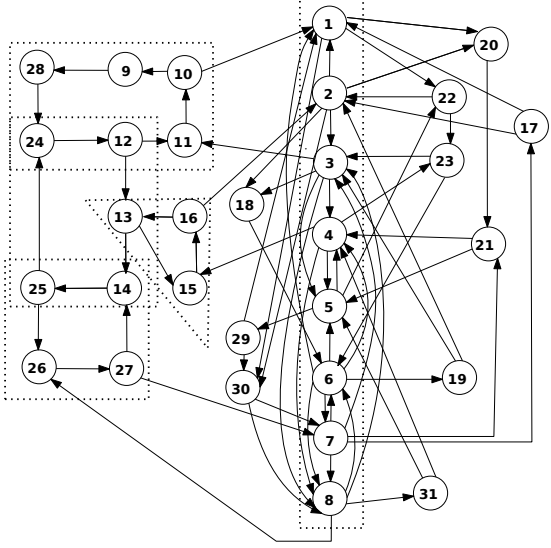}
			\caption{An IC structure $G_8$ which obeys neither Condition \ref{cond: ncyc1} nor \ref{cond: ncyc2}.}
			\label{n_cyc_example4}
		\end{figure}

		\begin{figure}[!t]
			\centering
			\includegraphics[width=17pc]{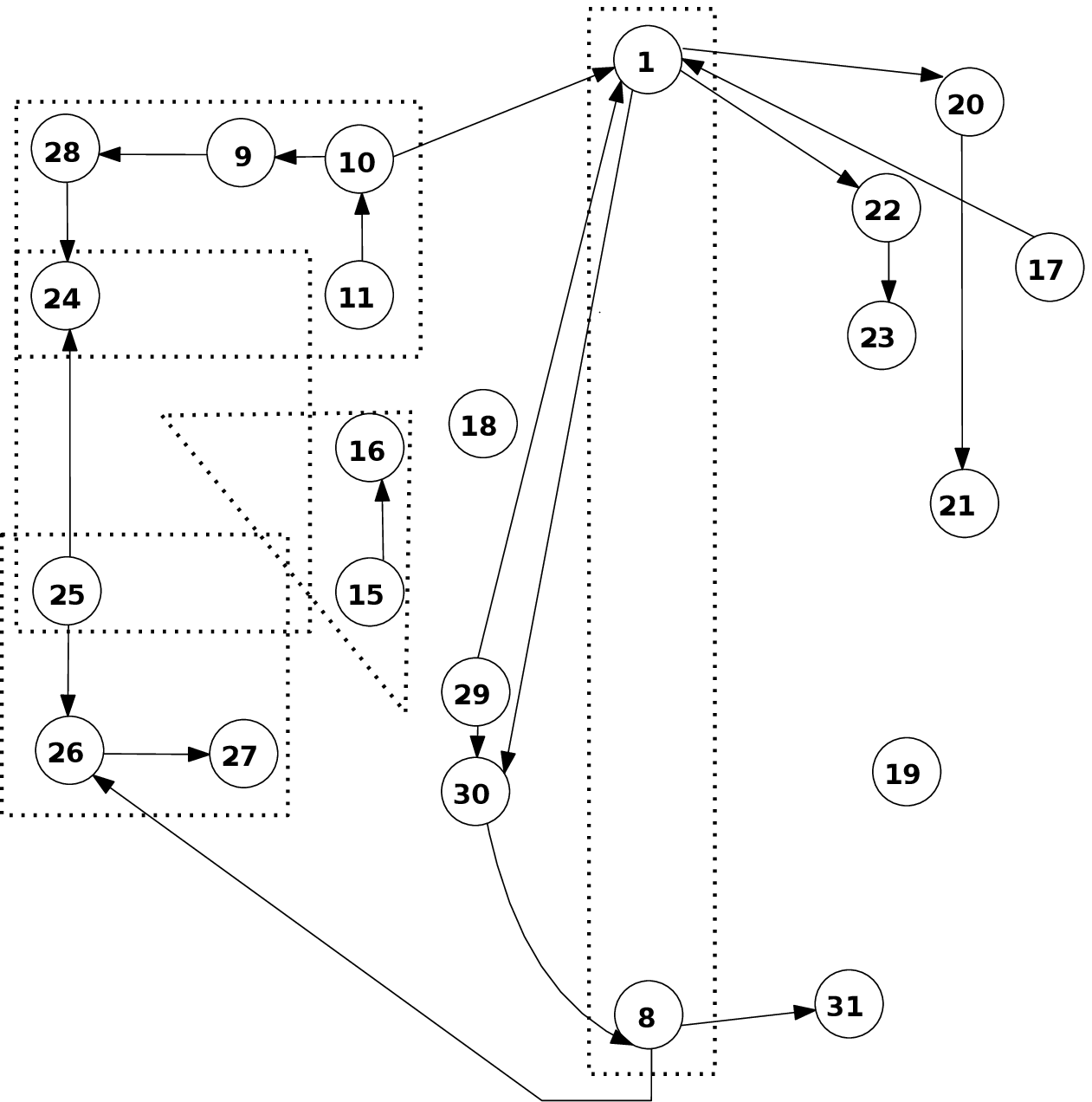}
			\caption{A maximum induced acyclic sub-digraph of the IC structure $G_8$}
			\label{n_cyc_example4_acyclic}
		\end{figure}
		
		\begin{figure}[!t]
			\centering
			\includegraphics[width=17pc]{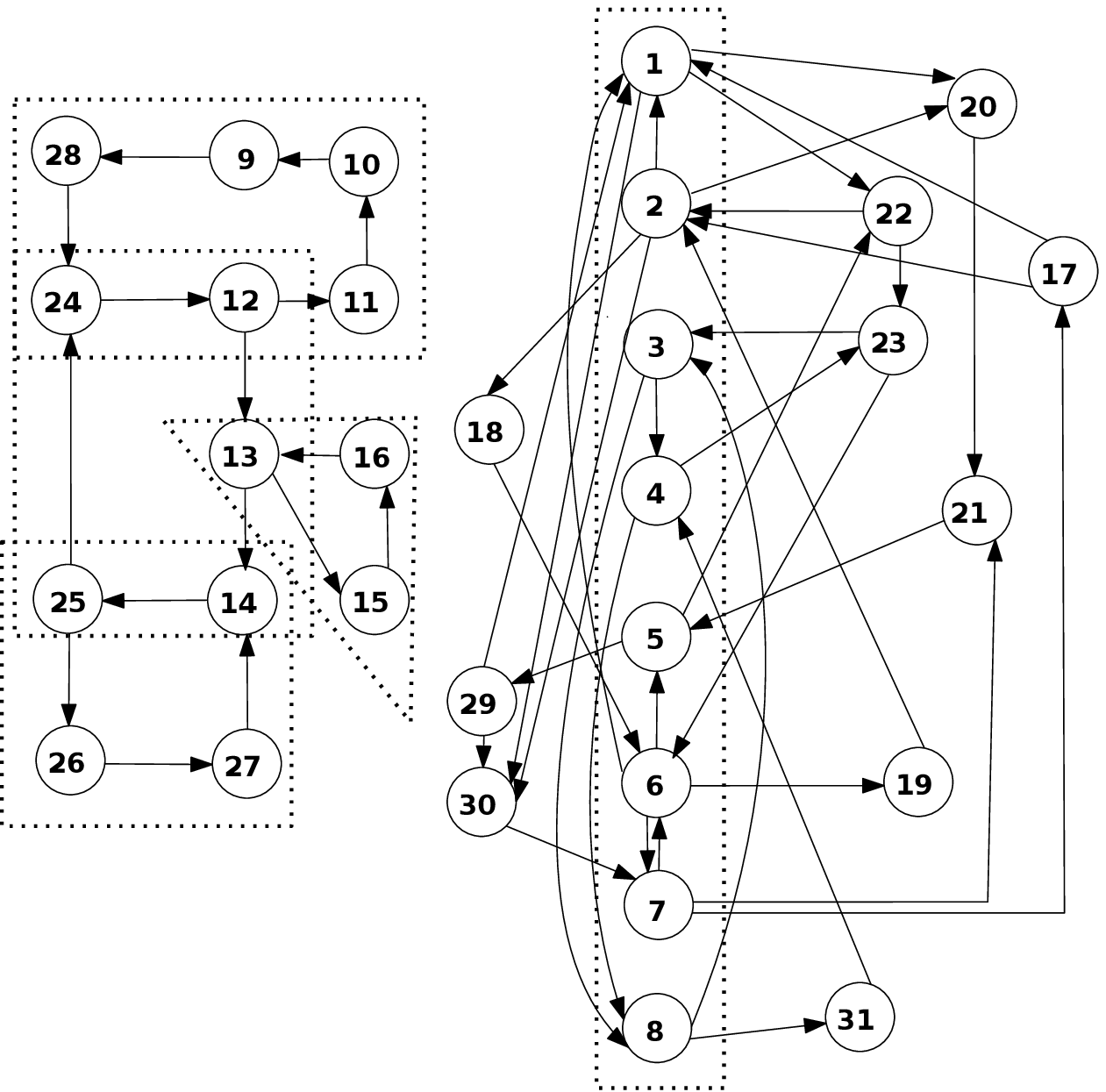}
			\caption{IC structures $G_8^{1}$ and $G_8^{2}$ with inner vertex sets $\{1,2,5,6,7\}$ and $\{3,4,8\}$ respectively.}
			\label{n_cyc_example4_VC1}
		\end{figure}
		
		\begin{figure}[!t]
			\centering
			\includegraphics[width=17pc]{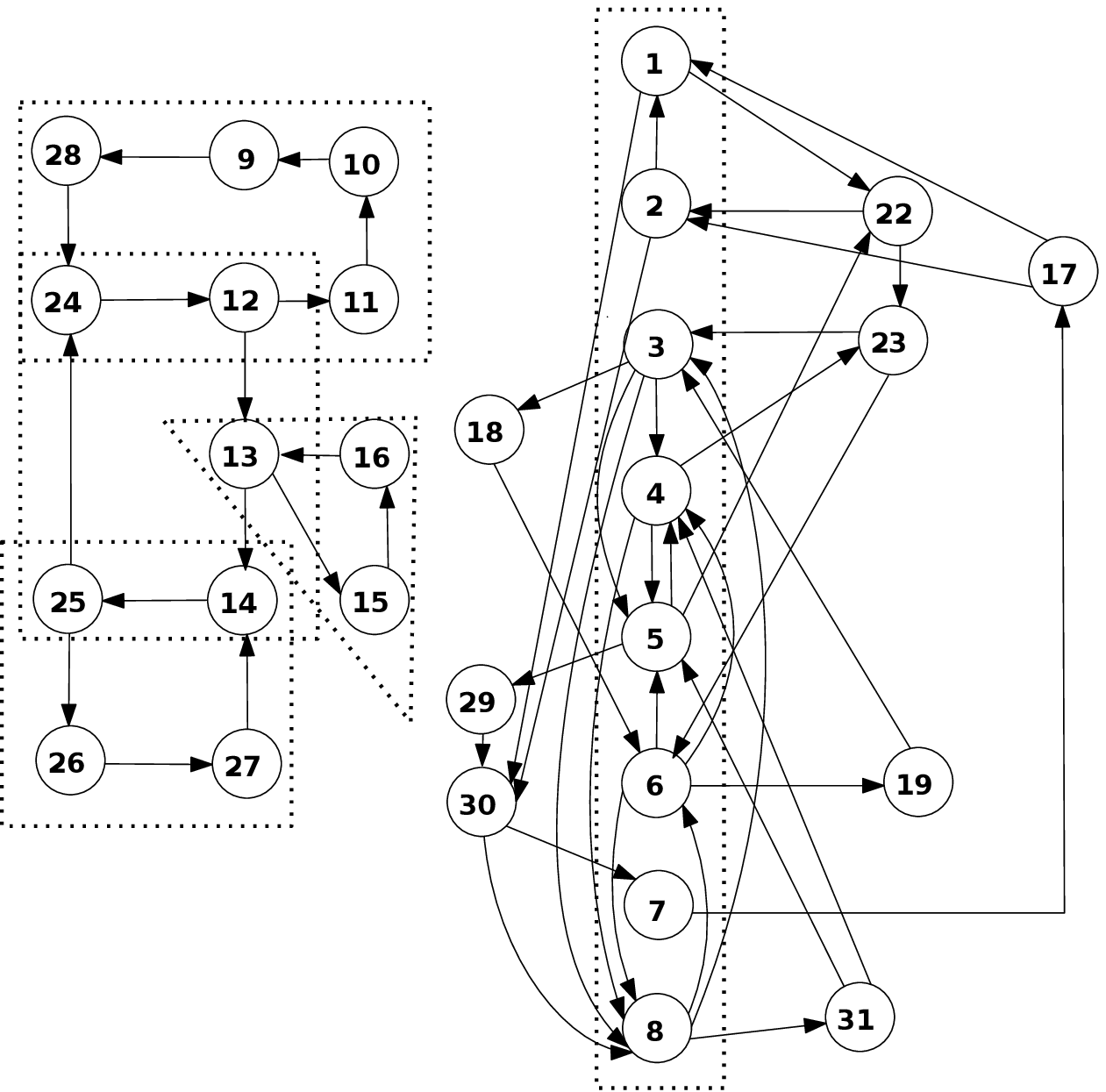}
			\caption{IC structures $G_8^{1}$ and $G_8^{2}$ with inner vertex sets $\{1,2,7\}$ and $\{3,4,5,6,8\}$ respectively.}
			\label{n_cyc_example4_VC2}
		\end{figure}
		
		\begin{figure}[!t]
			\centering
			\includegraphics[width=15pc]{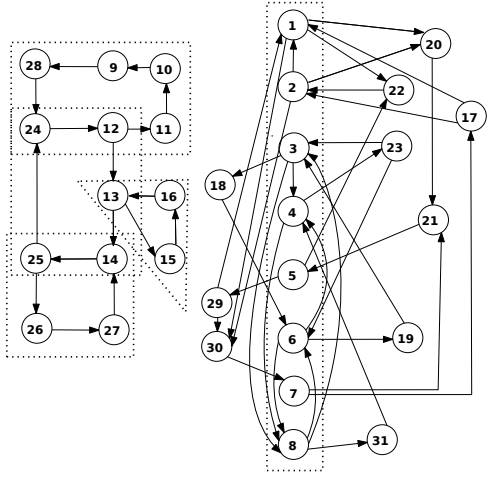}
			\caption{ IC structures $G_8^{1}$ and $G_8^{2}$ with inner vertex sets $\{1,2,5,7\}$ and $\{3,4,6,8\}$, and the cycles $C_1,C_2,C_3$ and $C_4$ in the IC structure $G_8$.}
			\label{n_cyc_example4_G1G2}
		\end{figure}
		
		\begin{exmp}
			Consider an IC structure $G_8$ shown in the Fig. \ref{n_cyc_example4}. For this IC structure $N=31, K=8,V_I=\{1,2,3,4,5,6,7,8\}$ and $V_{NI}=\{9,10,...,31\}$. Interlocked outer cycles are $C_1,C_2,C_3$ and $C_4$ with $V_{C_1}=
			\{12,13,14,25,24\},V_{C_2}=\{13,15,16\},V_{C_3}=\{11,10,9,28,24,12\}, V_{C_4} = \{25,26,27,14\} , V_{1,2}=\{13\}, V_{1,3}=\{24,12\},V_{1,4}= \{14,25\} $ and $ V_{2,3}=V_{2,4}=V_{3,4}= \phi $. There is no vertex in common with all the four cycles. $V_I^{in}=\{1,2,7\}, V_I^{out} = \{3,4,8\}$ and $V_I^{*} = \{5,6\}$. The induced sub-digraph with vertices $V_{NI} \cup \{1,8\} \backslash \{12,13,14\}$ is acyclic (shown in Fig. \ref{n_cyc_example4_acyclic}). Therefore $MAIS(G_K) \geq 22$ and hence $\beta(G_K) \geq 22$. This IC structure $G_8$ obeys neither Condition \ref{cond: ncyc1}, since the I-path from $4$ to $3$ and the I-path from $5$ to $6$ have the vertex $23$ in common (shown in Fig. \ref{n_cyc_example4_VC1}), nor \ref{cond: ncyc2}, since the I-path from $1$ to $2$ and the I-path from $5$ to $6$ have the vertex $22$ in common (shown in Fig. \ref{n_cyc_example4_VC2}).
			
			Let $V_I^{1}=\{1,2,5,7\}$ and $V_I^{2}=\{3,4,6,8\}$. Let $G_8^{1}$ be the induced sub-digraph of $G_8$ formed with the vertices $V_I^{1}$ and all the vertices in the I-path from each vertex in $V_I^{1}$ to every other vertex in $V_I^{1}$ and $G_8^{2}$ be the induced sub-digraph of $G_8$ formed with the vertices $V_I^{2}$ and all the vertices in the I-path from each vertex in $V_I^{2}$ to every other vertex in $V_I^{2}$. IC structures $G_8^{1}$ and $G_8^{2}$ are disjoint and they are disjoint from $C_1,C_2,C_3$ and $C_4$ (shown in Fig. \ref{n_cyc_example4_G1G2}). A set of maximum number of disjoint cycles in $\cal{C}$ is $C^{S}=\{C_2,C_3,C_4\}$. IC structures $G_8^{1},G_8^{2}$ and the cycles $C_2,C_3$ and $C_4$ are disjoint from one another.\\ $V_{NI}^{1}= \{20,21,22,17,29,30\},\\V_{NI}^{2}= \{23,18,19,31\},\\ V^{1}=\{1,2,5,7,20,21,22,17,29,30\},\\ V^{2}=\{3,4,6,8,23,18,19,31\}$ and\\ $V \backslash ((\cup_{C_k \in C^{S}}V_{C_k}) \cup V^{1} \cup V^{2})=\phi $. \\Using Algorithm \ref{algo2}, 
			\begin{enumerate}
				\item  $w_I^1 = x_1 \oplus x_2 \oplus x_5 \oplus x_7.$
				
				\item $ w_I^2 = x_3 \oplus x_4 \oplus x_6 \oplus x_8.$
				
				\item $V_{NI}^{1} \cup V_{NI}^{2}= \{20,21,22,17,29,30,23,18,19,31\}$.\\ $N_{G_K^{1}}^{+}(20)=\{21\},\\N_{G_K^{1}}^{+}(21)=\{5\},\\N_{G_K^{1}}^{+}(22)=\{2\},\\N_{G_K^{1}}^{+}(17)=\{1,2\},\\N_{G_K^{1}}^{+}(29)=\{30,1\}$,\\ $N_{G_K^{1}}^{+}(30)=\{7\}$,\\ $N_{G_K^{2}}^{+}(23)=\{3,6\},\\ N_{G_K^{2}}^{+}(18)=\{6\},\\ N_{G_K^{2}}^{+}(19)=\{3\}$ and\\ $N_{G_K^{2}}^{+}(31)=\{4\}$.\\ $w_{20}=x_{20} \oplus x_{21},\\ w_{21}= x_{21} \oplus x_5 ,\\ w_{22}=x_{22} \oplus x_2,\\ w_{17}= x_{17} \oplus x_1 \oplus x_2,\\w_{29}=x_{29} \oplus x_{30} \oplus x_1,\\w_{30}=x_{30} \oplus x_7,\\ w_{23}=x_{23} \oplus x_6 \oplus x_3,\\ w_{18}=x_{18} \oplus x_6, \\w_{19}=x_{19} \oplus x_3 $ and\\ $w_{31}=x_{31} \oplus x_4.$
				
				\item For $C_2 \in C^{S}$, an index code $\{  x_{13} \oplus x_{15},x_{15} \oplus x_{16} \}$ of length $2$ is obtained.
				
				\item For $C_3 \in C^{S}$, an index code $\{x_{28} \oplus x_{24} , x_{9} \oplus x_{28},x_{10} \oplus x_9, x_{11} \oplus x_{10}, x_{12} \oplus x_{11}\}$ of length $5$ is obtained.
				
				\item For $C_4 \in C^{S}$, an index code $\{ x_{25} \oplus x_{26},x_{26} \oplus x_{27} , x_{27} \oplus x_{14} \}$ of length $3$ is obtained.			
				
			\end{enumerate}
			
			Number of transmitted symbols is $22$, which achieves the lower bound on the optimal broadcast rate. Hence this code is optimal.
			
			Although this IC structure obeys neither Condition \ref{cond: ncyc1} nor Condition \ref{cond: ncyc2}, we have shown that the code construction given by Algorithm \ref{algo2} is still optimal for this example. \label{exmp: VC1VC2_optimal}
		\end{exmp}

	\begin{figure}[!t]
		\centering
		\includegraphics[width=21pc]{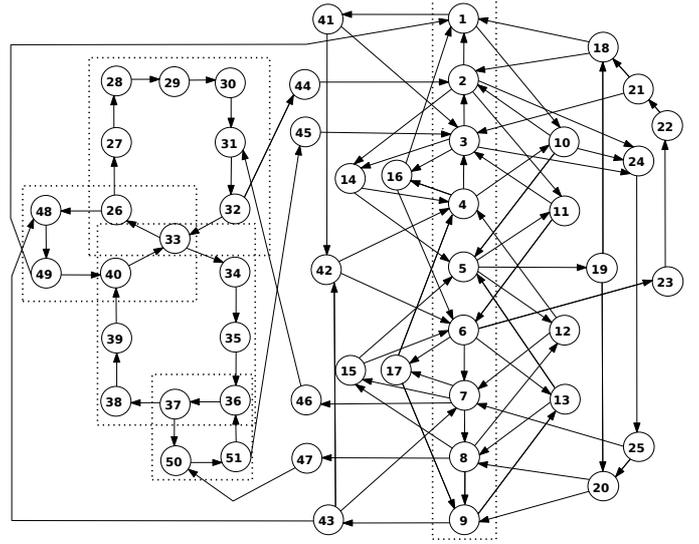}
		\caption{An IC structure $G_9$ which obeys neither Condition \ref{cond: ncyc1} nor \ref{cond: ncyc2}.}
		\label{n_cyc_example3}
	\end{figure}
	
	\begin{figure}[!t]
		\centering
		\includegraphics[width=17pc]{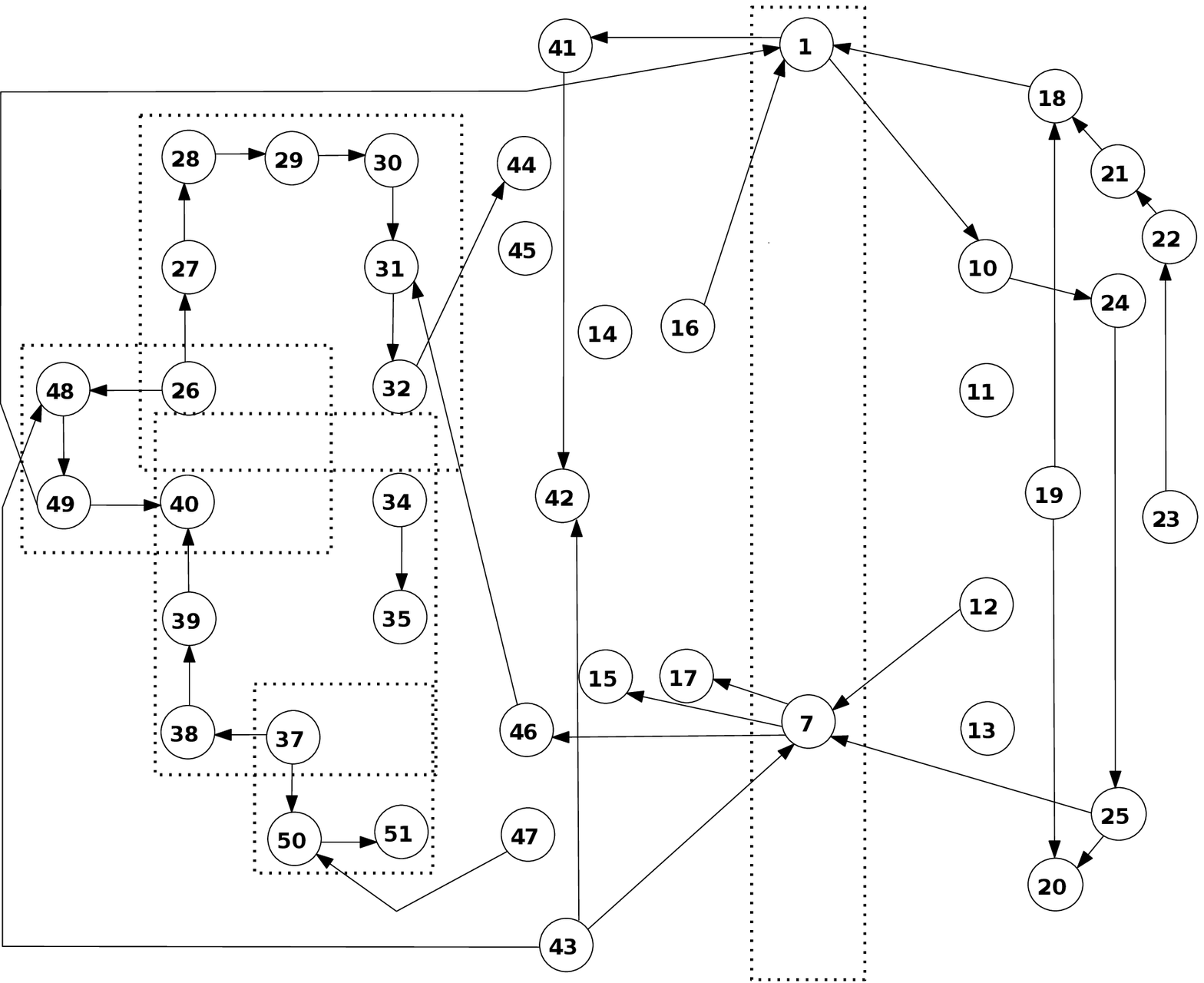}
		\caption{A maximum induced acyclic sub-digraph of the IC structure $G_9$}
		\label{n_cyc_example3_acyclic}
	\end{figure}
	
	\begin{figure}[!t]
		\centering
		\includegraphics[width=20pc]{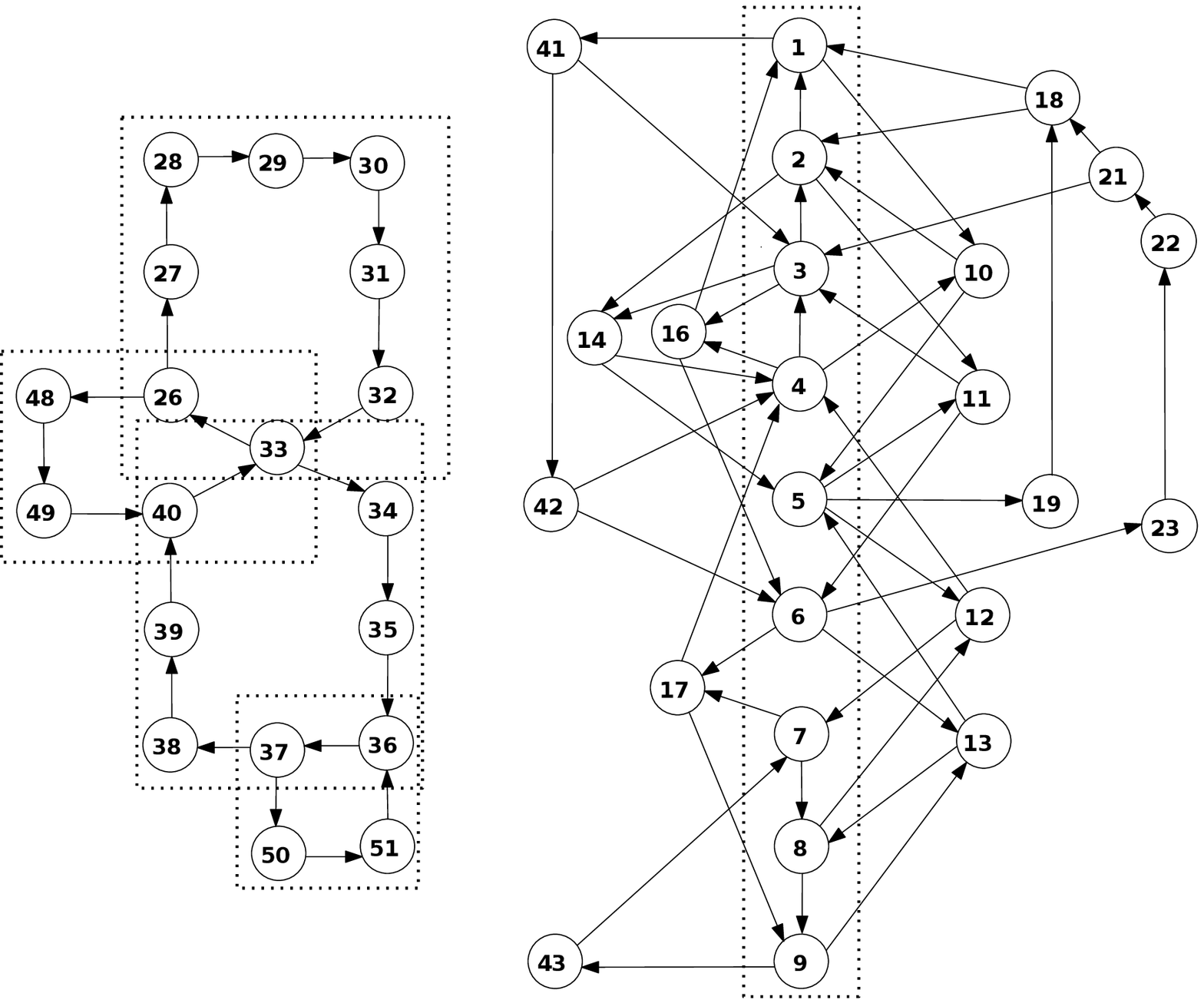}
		\caption{$C_1,C_2,C_3,C_4$ and IC structures $G_9^{1}$ and $G_9^{2}$ with inner vertex sets $\{1,2,3,4,5,6\}$ and $\{7,8,9\}$ respectively.}
		\label{n_cyc_example3_VC1}
	\end{figure}
	
	\begin{figure}[!t]
		\centering
		\includegraphics[width=17pc]{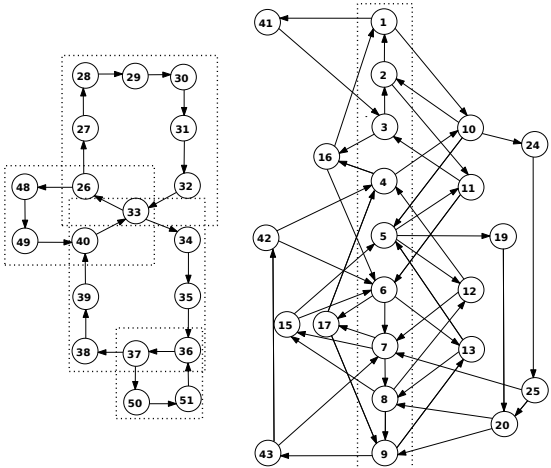}
		\caption{$C_1,C_2,C_3,C_4$ and IC structures $G_9^{1}$ and $G_9^{2}$ with inner vertex sets $\{1,2,3\}$ and $\{4,5,6,7,8,9\}$ respectively.}
		\label{n_cyc_example3_VC2}
	\end{figure}
	
	\begin{exmp}
		Consider the IC structure $G_{9}$ shown in the Fig. \ref{n_cyc_example3}. For this IC structure $N=51, K=9, V_I=\{1,2,3,4,5,6,7,8,9\}$ and $V_{NI}=\{10,11,12...,51\}$. Interlocked outer cycles are $C_1,C_2,C_3$ and $C_4$ with $V_{C_1}=
		\{26,27,28,29,30,31,32,33\},V_{C_2}=\{33,34,35,36,37,38,39,40\},V_{C_3}=\{26,48,49,40,33\}, V_{C_4}=\{36,37,50,51\} , V_{1,2}=\{33\}, V_{1,3}=\{33,26\},V_{2,3}=\{40,33\},V_{2,4}=\{36,37\}$ and $V_{1,4}=V_{3,4}= \phi$. $V_I^{in} = \{1,2,3\}$, $V_I^{out} = \{7,8,9\}$ and $V_I^{*} = \{4,5,6\}$. The induced sub-digraph with vertices $V_{NI} \cup \{1,7\} \backslash \{33,36\}$ is acyclic (shown in Fig. \ref{n_cyc_example3_acyclic}). Therefore $MAIS(G_K) \geq 42$ and hence $\beta(G_K) \geq 42$. 
		
		This IC structure obeys neither Condition \ref{cond: ncyc1}, since the I-path from $8$ to $7$ and the I-path from $5$ to $4$ have the vertex $12$ in common (shown in Fig. \ref{n_cyc_example3_VC1}), nor  \ref{cond: ncyc2}, since the I-path from $1$ to $2$ and the I-path from $4$ to $5$ have the vertex $10$ in common (shown in Fig. \ref{n_cyc_example3_VC2}).
		
		We will show that we cannot partition the inner vertex set $V_I$ into two sets such that, both the sets form two disjoint IC structures and these two IC structures are disjoint from $C_1,C_2,C_3$ and $C_4$.
		
		The inner vertex set $V_I$ can be partitioned into two disjoint non-empty subsets in $510$ $(\binom{9}{1}+\binom{9}{2}+\binom{9}{3}+\binom{9}{4}+\binom{9}{5}+\binom{9}{6}+\binom{9}{7}+\binom{9}{8})$ ways. Out of the $510$ partitions possible, $502$ partitions include one or more vertices from $V_I^{in}$ and one or more vertices from $V_I^{out}$ in the same set. If some vertex $i$ from $V_I^{in}$ and some vertex $j$ from $V_I^{out}$ are in the same set, then that set doesn't form an IC structure which is disjoint from the cycles $C_1,C_2,C_3$ and $C_4$, since the I-path from $j$ to $i$ passes through the cycles. Hence those $502$ partitions are ruled out. Then there are $8$ partitions remaining.  
		
		If the vertices $4$ and $5$ are included in $V_I^{in}$, the I-path from $5$ to $4$ intersects the I-path from $8$ to $7$ and if the vertices $4$ and $5$ are included in $V_I^{out}$, the I-path from $4$ to $5$ intersects the I-path from $1$ to $2$. Hence the vertices $4$ and $5$ cannot be included in the same set. Therefore, out of the remaining $8$ partitions $4$ are ruled out. 
		
		If the vertex $4$ is included in $V_I^{in}$ and the vertex $5$ in $V_I^{out}$, the vertex $6$ cannot be included in any of the subsets $V_I^{out}$ or $V_I^{in}$. If we include it in $V_I^{in}$, the I-path from $6$ to $4$ intersects the I-path from $7$ to $9$ and if we include it in $V_I^{out}$, the I-path from $5$ to $6$ intersects the I-path from $2$ to $3$. Hence out of the remaining $4$ partitions $2$ are ruled out.
		
		If the vertex $4$ is included in $V_I^{out}$ and the vertex $5$ in $V_I^{in}$, the vertex $6$ cannot be included in any of the subsets $V_I^{out}$ or $V_I^{in}$. If we include it in $V_I^{in}$, the I-path from $6$ to $5$ intersects the I-path from $9$ to $8$ and if we include it in $V_I^{out}$, the I-path from $4$ to $6$ intersects the I-path from $3$ to $1$. Hence the remaining $2$ partitions are also ruled out.
		
		Hence we cannot partition the inner vertex set $V_I$ into two sets, such that both the sets form two disjoint IC structures and these two IC structures are disjoint from the cycles $C_1,C_2,C_3$ and $C_4$. 
		
		This IC structure obeys neither Condition \ref{cond: ncyc1} nor \ref{cond: ncyc2} and the code construction given by Algorithm \ref{algo2} is not valid for this example.
	\end{exmp}

	\begin{figure}[!t]
		\centering
		\includegraphics[width=17pc]{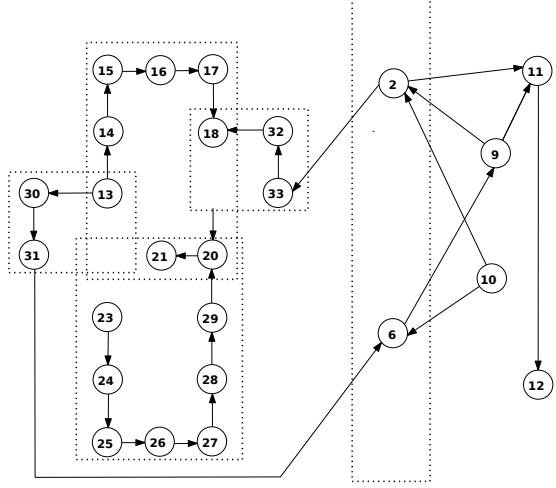}
		\caption{A maximum induced acyclic sub-digraph of the IC structure $G_8$}
		\label{n_cyc_example1_acyclic}
	\end{figure}
		
	\begin{figure}[!t]
		\centering
		\includegraphics[width=17pc]{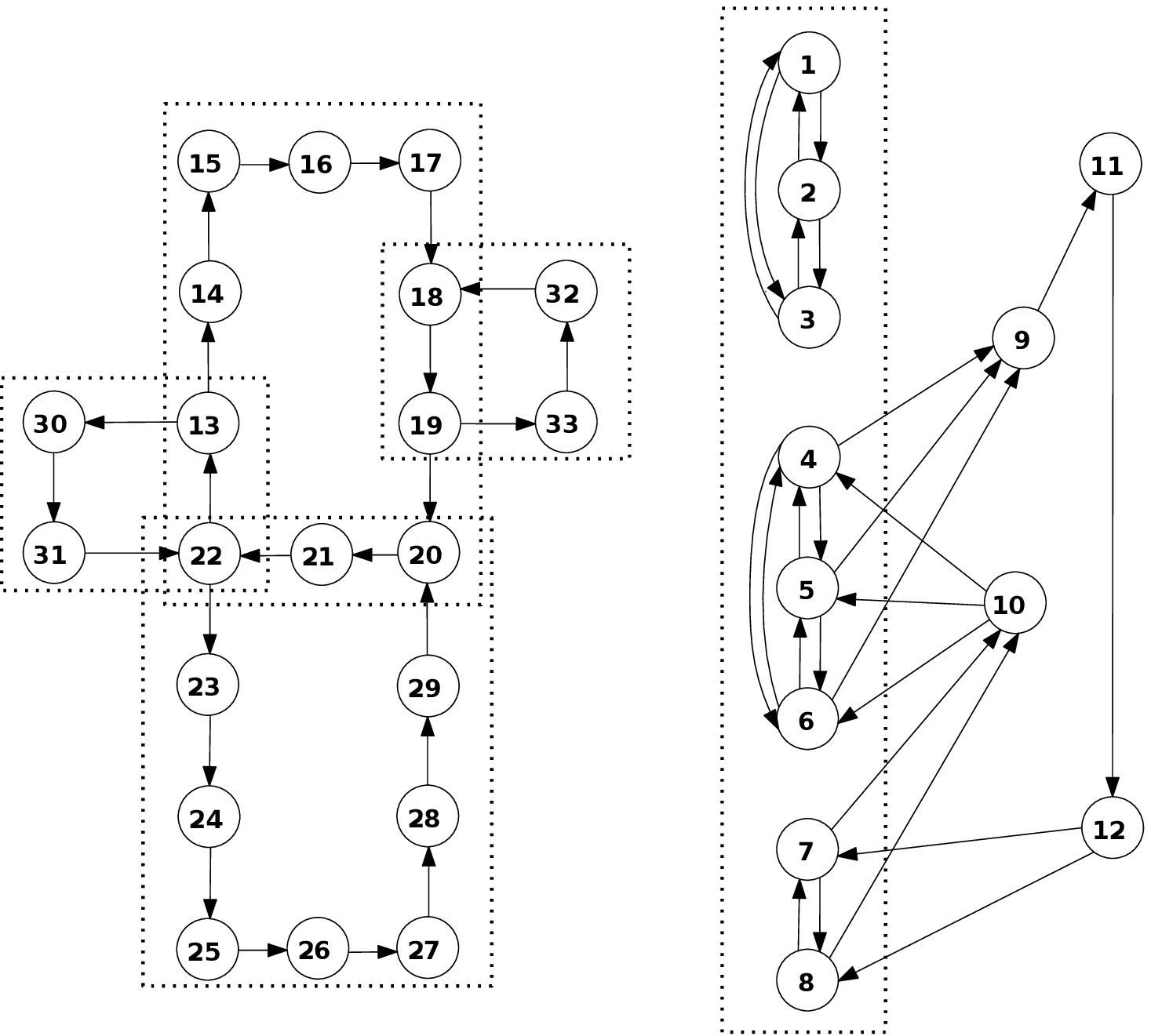}
		\caption{$C_1,C_2,C_3,C_4$ and IC structures $G_8^{1}$ and $G_8^{2}$ with inner vertex sets $\{1,2,3\}$ and $\{4,5,6,7,8\}$ respectively.}
		\label{n_cyc_example1_C1}		
	\end{figure}
	
	\begin{figure}[!t]
		\centering
		\includegraphics[width=17pc]{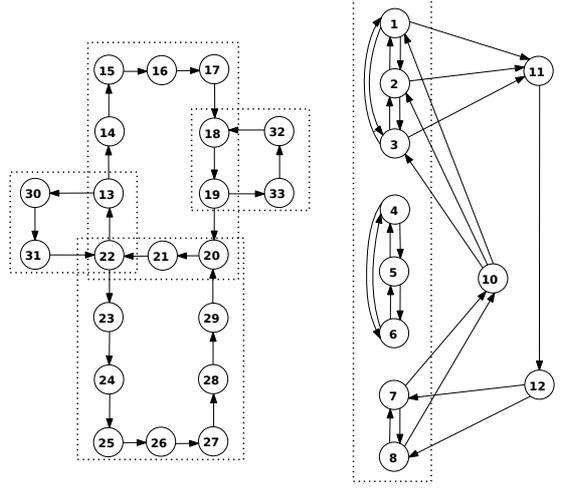}
		\caption{$C_1,C_2,C_3,C_4$ and IC structures $G_8^{1}$ and $G_8^{2}$ with inner vertex sets $\{1,2,3,7,8\}$ and $\{4,5,6\}$ respectively.}
		\label{n_cyc_example1_C2}
	\end{figure}

	\begin{exmp}		
		Consider Example \ref{exm: 5_code}. The induced sub-digraph with vertices $V_{NI} \cup \{2,6\} \backslash \{22,19\}$ is acyclic (shown in Fig. \ref{n_cyc_example1_acyclic}). Therefore $MAIS(G_K) \geq 25$ and hence $\beta(G_K) \geq 25$. This IC structure obeys both Condition \ref{cond: ncyc1} (shown in Fig. \ref{n_cyc_example1_C1}) and Condition \ref{cond: ncyc2} (shown in Fig. \ref{n_cyc_example1_C2}).

		 Number of transmitted symbols is $25$, which achieves the lower bound on the optimal broadcast rate. Hence this code is optimal.
		
		This IC structure obeys both Condition \ref{cond: ncyc1} and \ref{cond: ncyc2} and we have shown that the index code constructed using Algorithm \ref{algo2} is optimal for this example. 			
	\end{exmp}

		\begin{figure}[!t]
			\centering
			\includegraphics[width=17pc]{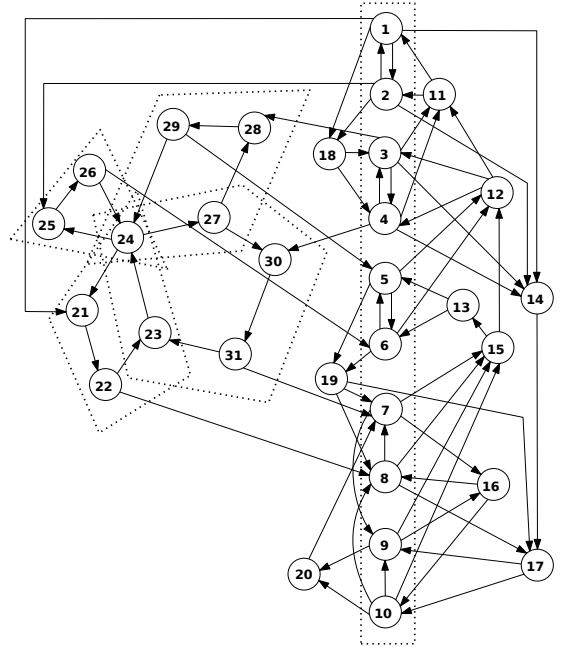}
			\caption{An IC structure $G_{10}$ which obeys Condition \ref{cond: ncyc1}. }
			\label{n_cyc_example2}
		\end{figure}

		\begin{figure}[!t]
			\centering
			\includegraphics[width=17pc]{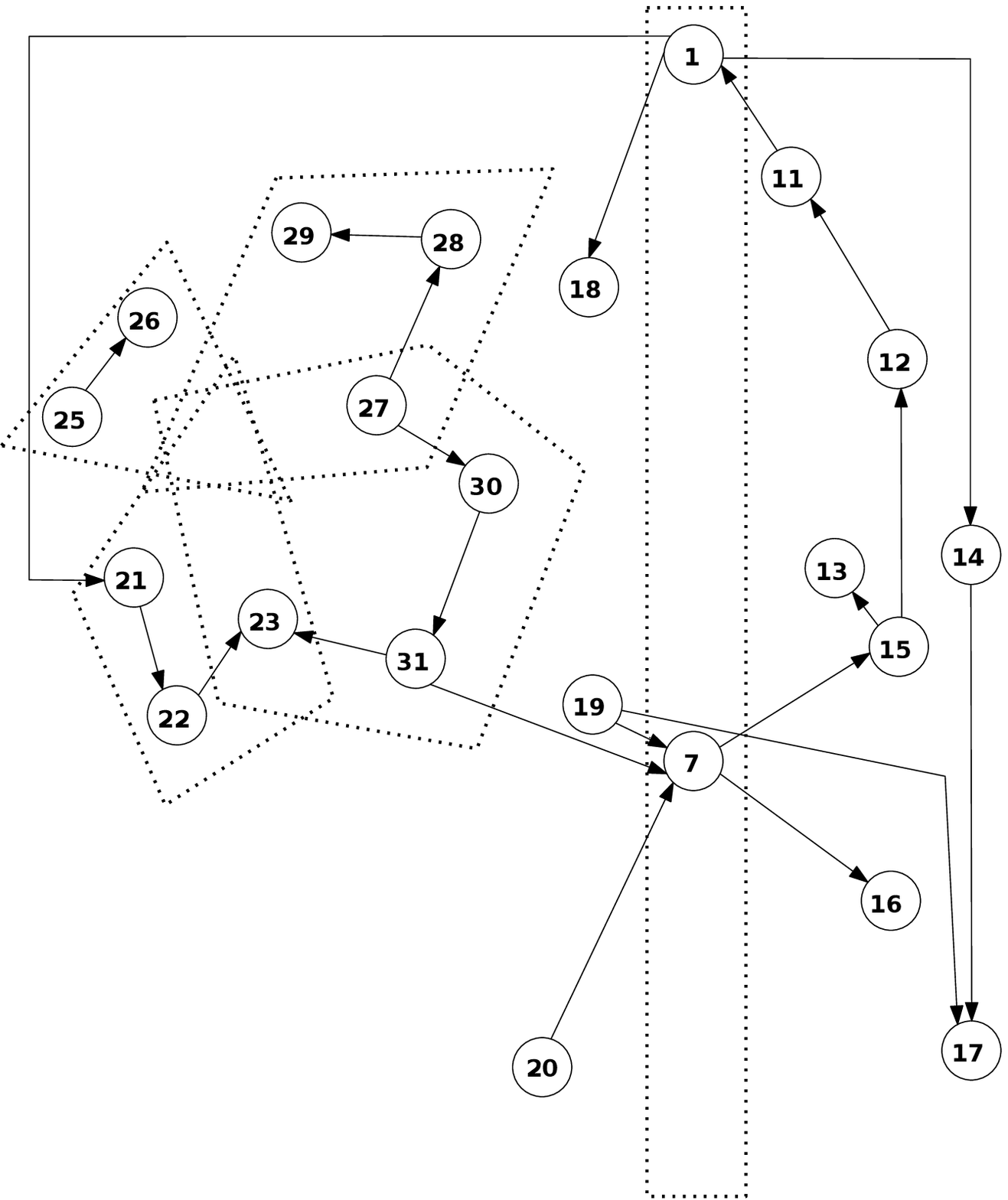}
			\caption{A maximum induced acyclic sub-digraph of the IC structure $G_{10}$}
			\label{n_cyc_example2_acyclic}
		\end{figure}
		
		\begin{figure}[!t]
			\centering
			\includegraphics[width=17pc]{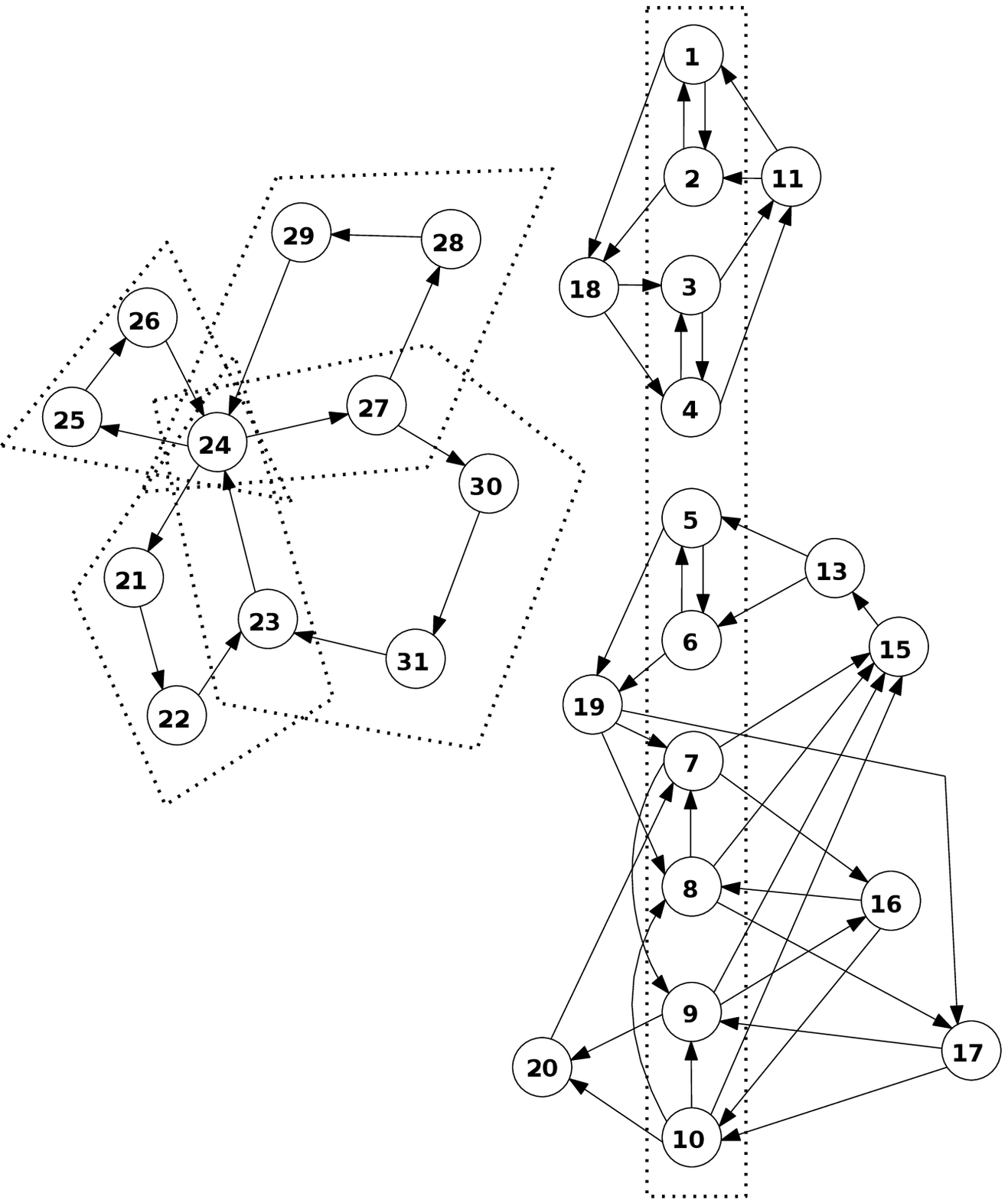}
			\caption{$C_1,C_2,C_3,C_4$ and IC structures $G_{10}^{1}$ and $G_{10}^{2}$ with inner vertex sets $\{1,2,3,4\}$ and $\{5,6,7,8,9,10\}$ respectively.}
			\label{n_cyc_example2_C1}			
		\end{figure}
		
		\begin{figure}[!t]
			\centering
			\includegraphics[width=17pc]{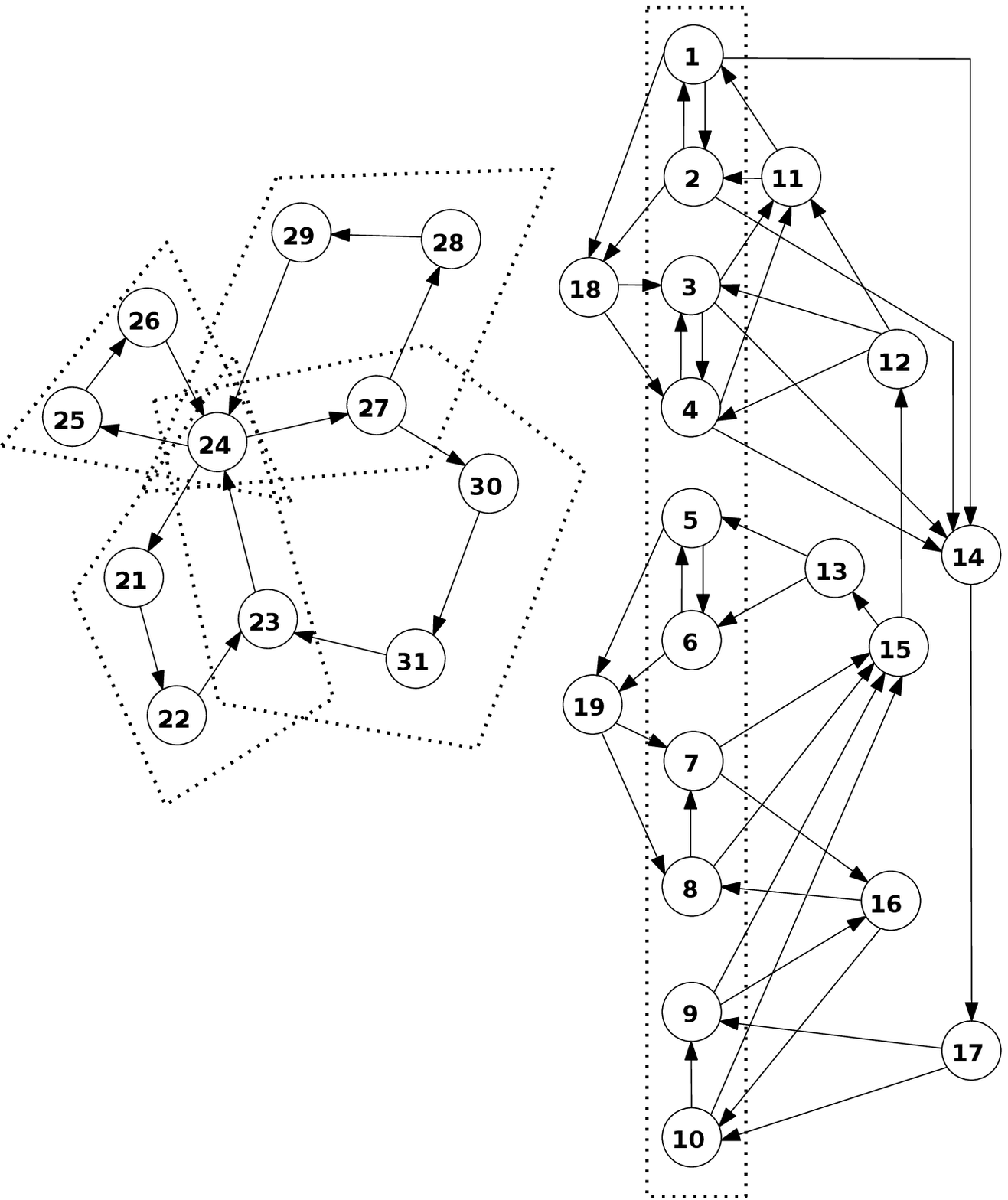}
			\caption{$C_1,C_2,C_3,C_4$ and IC structures $G_{10}^{1}$ and $G_{10}^{2}$ with inner vertex sets $\{1,2,3,4,9,10\}$ and $\{5,6,7,8\}$ respectively.}
			\label{n_cyc_example2_VC2}
		\end{figure}

		\begin{exmp}
		Consider an IC structure $G_{10}$ shown in the Fig. \ref{n_cyc_example2}. For this IC structure $N=31, K=10, V_I=\{1,2,3,...,9,10\}$ and $V_{NI}=\{11,12,...,31\}$. Interlocked outer cycles are $C_1,C_2,C_3$ and $C_4$ with $V_{C_1}=
		\{24,25,26\},V_{C_2}=\{27,28,29,24\},V_{C_3}=\{23,24,27,30,31\},V_{C_4}=\{21,22,23,24\}, V_{1,2}=V_{1,3}=V_{1,4}=V_{2,4}=\{24\}, V_{2,3}=\{24,27\} $ and $V_{3,4}=\{23,24\}$. For this IC structure $V_I^{in} = \{5,6,7,8\}$, $V_I^{out} = \{1,2,3,4\}$ and $V_I^{*} = \{9,10\}$. The induced sub-digraph with vertices $V_{NI} \cup \{1,7\} \backslash \{24\}$ is acyclic (shown in Fig. \ref{n_cyc_example2_acyclic}). Therefore $MAIS(G_K) \geq 22$ and hence $\beta(G_K) \geq 22$. This IC structure obeys Condition \ref{cond: ncyc1} (shown in Fig. \ref{n_cyc_example2_C1}) and doesn't obey Condition \ref{cond: ncyc2} since the I-path from $9$ to $1$ and the I-path from $7$ to $5$ have the vertex $15$ in common (shown in Fig. \ref{n_cyc_example2_VC2}). Hence $V_I^1=\{5,6,7,8,9,10\}$ and $V_I^2=\{1,2,3,4\}$. 
		
		Let $G_{10}^{1}$ be the induced sub-digraph of $G_{10}$ formed with the vertices $V_I^{1}$ and all the vertices in the I-path from each vertex in $V_I^{1}$ to every other vertex in $V_I^{1}$ and $G_{10}^{2}$ be the induced sub-digraph of $G_{10}$ formed with the vertices $V_I^{2}$ and all the vertices in the I-path from each vertex in $V_I^{2}$ to every other vertex in $V_I^{2}$. A set of maximum number of disjoint cycles in $\cal{C}$ is $C^{S}=\{C_1\}$.
		\\
		$V_{NI}^{1}= \{18,11\},\\V_{NI}^{2}= \{19,20,13,15,16,17\},\\ V^{1}=\{1,2,3,4,18,11\},\\ V^{2}=\{5,6,7,8,9,10,19,20,13,15,16,17\}$ and \\$V \backslash ((\cup_{C_k \in C^{S}}V_{C_k}) \cup V^{1} \cup V^{2})=\{  14,12,21,22,23,27,28,29,30,31\} $.\\ Using Algorithm \ref{algo2}, 
			\begin{enumerate}
				\item  $w_I^1 = x_1 \oplus x_2 \oplus x_3 \oplus x_4.$
				
				\item $ w_I^2 = x_5 \oplus x_6 \oplus x_7 \oplus x_8 \oplus x_9 \oplus x_{10}.$
				
				\item $V_{NI}^{1} \cup V_{NI}^{2}= \{18,11,19,20,13,15,16,17\}$.\\ $N_{G_K^{1}}^{+}(18)=\{3,4\},$\\ $N_{G_K^{1}}^{+}(11)=\{1,2\}$,\\ $N_{G_K^{2}}^{+}(19)=\{7,8\}, \\N_{G_K^{2}}^{+}(20)=\{7\},\\ N_{G_K^{2}}^{+}(13)=\{5,6\},\\ N_{G_K^{2}}^{+}(15)=\{13\},\\N_{G_K^{2}}^{+}(16)=\{10,8\}$ and \\$N_{G_K^{2}}^{+}(17)=\{9,10\}$.\\ $w_{18}=x_{18} \oplus x_3 \oplus x_4, \\w_{11}= x_{11} \oplus x_1 \oplus x_2 , \\w_{19}=x_{19} \oplus x_7 \oplus x_8,\\ w_{20}= x_{20} \oplus x_7,\\w_{13}=x_{13} \oplus x_5 \oplus x_6,\\w_{15}= x_{15} \oplus x_{13}, \\w_{16}= x_{16} \oplus x_{10} \oplus x_8$ and\\ $w_{17}=x_{17} \oplus x_9 \oplus x_{10}.$
				
				\item For $C_1 \in C^{S}$, an index code $\{  x_{24} \oplus x_{25},x_{25} \oplus x_{26} \}$ of length $2$ is obtained.
				
					\item For each $j \in \{14,12,21,22,23,27,28,29,30,31\},\\$ $w_j=x_j.$
			
			\end{enumerate}
						
		Number of transmitted symbols is $22$, which achieves the lower bound on the optimal broadcast rate. Hence this scheme is optimal.
		
		This IC structure obeys Condition \ref{cond: ncyc1} and we have shown that the the index code constructed using Algorithm \ref{algo2} is optimal for this example. 	\label{exm: n_cyc_example2}
	\end{exmp}

		\begin{figure}[!t]
			\centering
			\includegraphics[width=17pc]{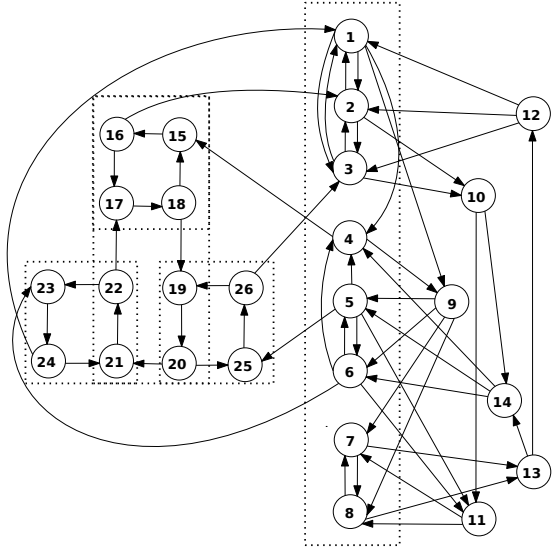}
			\caption{An IC structure $G_{8}$ which obeys Condition \ref{cond: ncyc2}. }
			\label{n_cyc_example5}
		\end{figure}

		\begin{figure}[!t]
			\centering
			\includegraphics[width=17pc]{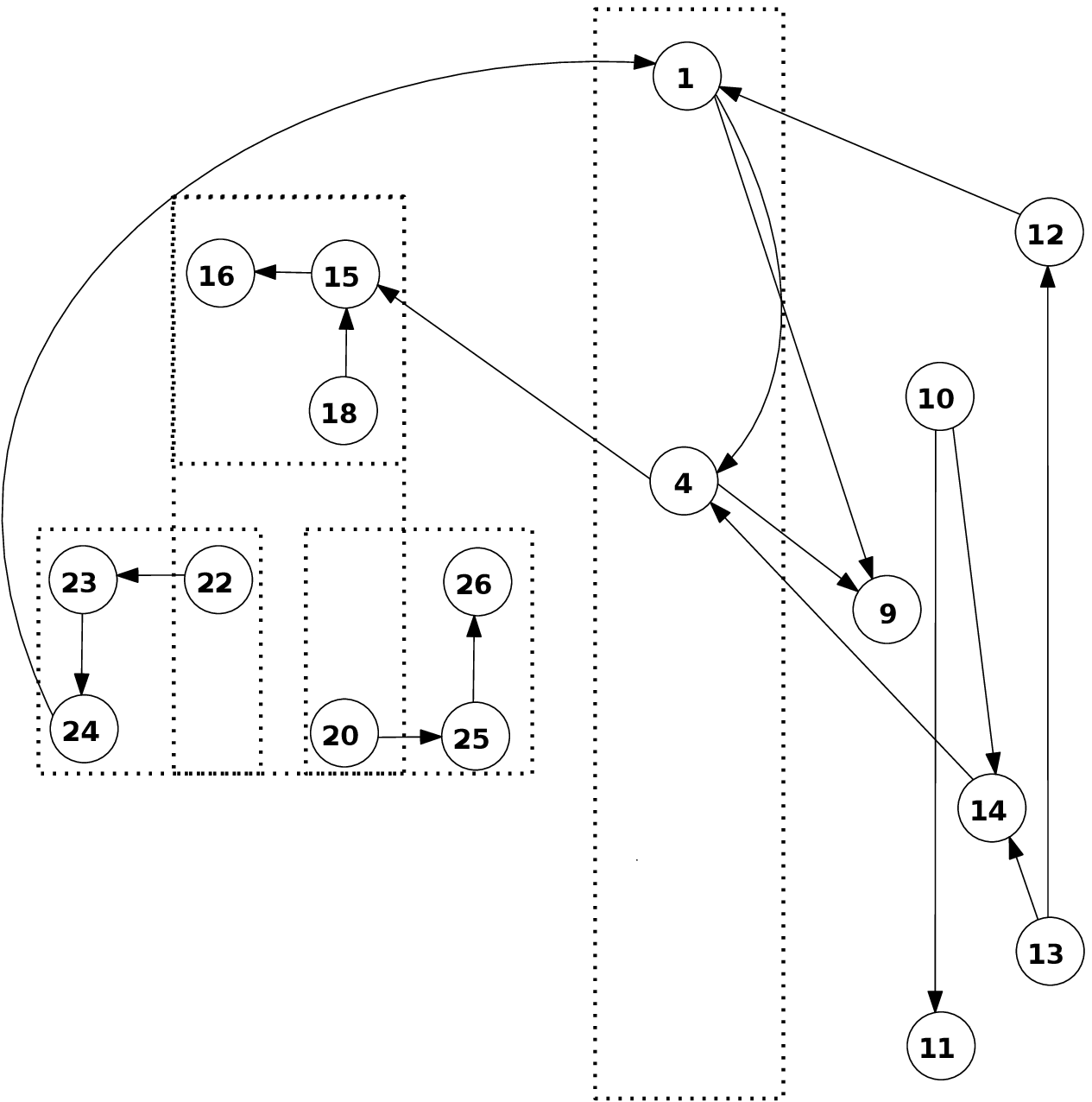}
			\caption{A maximum induced acyclic sub-digraph of the IC structure $G_{8}$}
			\label{n_cyc_example5_acyclic}
		\end{figure}
		
		\begin{figure}[!t]
			\centering
			\includegraphics[width=17pc]{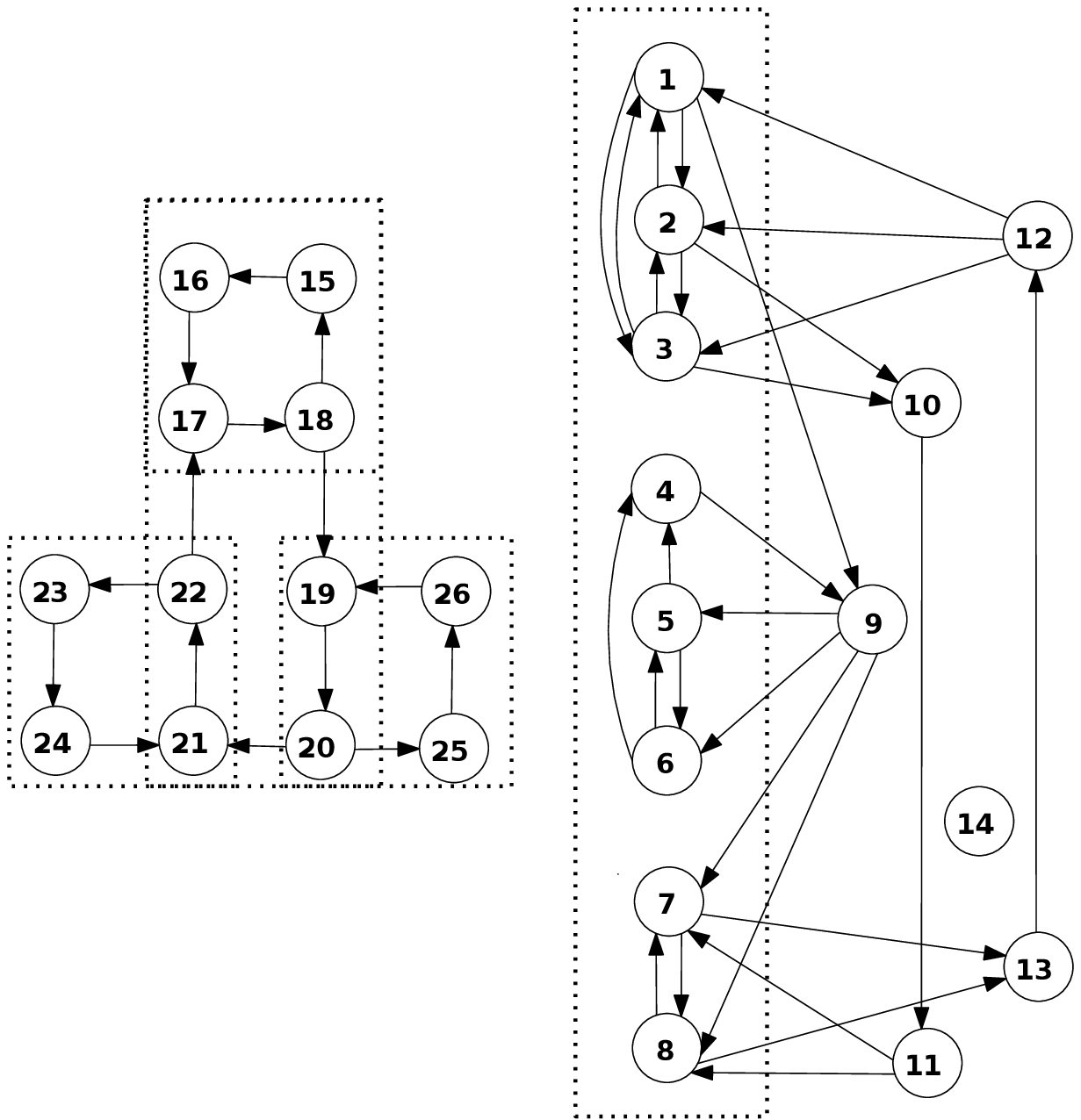}
			\caption{$C_1,C_2,C_3,C_4$ and IC structures $G_{8}^{1}$ and $G_{8}^{2}$ with inner vertex sets $\{4,5,6\}$ and $\{1,2,3,7,8\}$ respectively.}
			\label{n_cyc_example5_VC1}			
		\end{figure}
		
		\begin{figure}[!t]
			\centering
			\includegraphics[width=17pc]{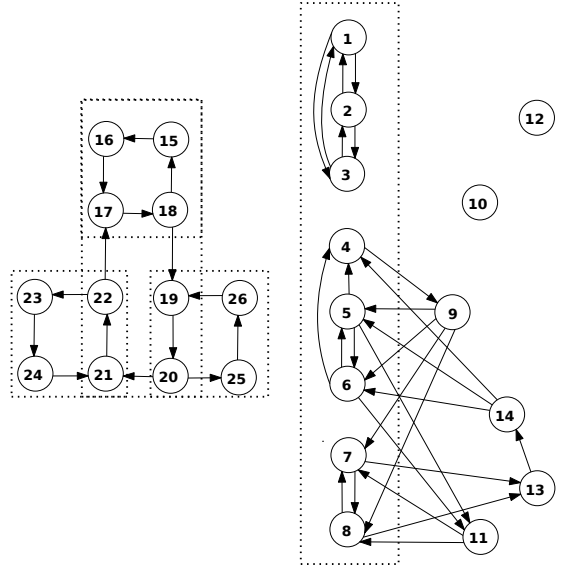}
			\caption{$C_1,C_2,C_3,C_4$ and IC structures $G_{8}^{1}$ and $G_{8}^{2}$ with inner vertex sets $\{4,5,6,7,8\}$ and $\{1,2,3\}$ respectively.}
			\label{n_cyc_example5_C2}
		\end{figure}
		
		\begin{exmp}
			Consider an IC structure $G_{8}$ shown in the Fig. \ref{n_cyc_example5}. For this IC structure $N=26, K=8, V_I=\{1,2,3,...,7,8\}$ and $V_{NI}=\{9,10,...,26\}$. Interlocked outer cycles are $C_1,C_2,C_3$ and $C_4$ with $V_{C_1}=
			\{17,18,19,20,21,22\},V_{C_2}=\{15,16,17,18\},V_{C_3}=\{21,22,23,24\},V_{C_4}=\{19,20,25,26\}, V_{1,2}=\{17,18\}, V_{1,3}= \{21,22\}, V_{1,4}= \{19,20\}$ and $V_{2,4}=V_{2,3}=V_{3,4}=\phi$. For this IC structure $V_I^{in} = \{1,2,3\}$, $V_I^{out} = \{4,5,6\}$ and $V_I^{*} = \{7,8\}$. The induced sub-digraph with vertices $V_{NI} \cup \{1,4\} \backslash \{17,19,21\}$ is acyclic (shown in Fig. \ref{n_cyc_example5_acyclic}). Therefore $MAIS(G_K) \geq 17$ and hence $\beta(G_K) \geq 17$. This IC structure doesn't obey Condition \ref{cond: ncyc1} since the I-path from $4$ to $5$ and the I-path from $1$ to $7$ have the vertex $9$ in common (shown in Fig. \ref{n_cyc_example5_VC1}) and obeys Condition \ref{cond: ncyc2} (shown in Fig. \ref{n_cyc_example5_C2}). Hence $V_I^1=\{1,2,3\}$ and $V_I^2=\{4,5,6,7,8\}$. 
			
			Let $G_8^{1}$ be the induced sub-digraph of $G_8$ formed with the vertices $V_I^{1}$ and all the vertices in the I-path from each vertex in $V_I^{1}$ to every other vertex in $V_I^{1}$ and $G_8^{2}$ be the induced sub-digraph of $G_8$ formed with the vertices $V_I^{2}$ and all the vertices in the I-path from each vertex in $V_I^{2}$ to every other vertex in $V_I^{2}$. A set of maximum number of disjoint cycles in $\cal{C}$ is $C^{S}=\{C_2,C_3,C_4\}$. 
				\\
				$V_{NI}^{1}= \phi,\\V_{NI}^{2}= \{9,14,13,11\},\\ V^{1}=\{1,2,3\},\\ V^{2}=\{4,5,6,7,8,9,14,13,11\}$ and \\$V \backslash ((\cup_{C_k \in C^{S}}V_{C_k}) \cup V^{1} \cup V^{2})=\{ 10,12\} $.\\ Using Algorithm \ref{algo2}, 
				\begin{enumerate}
					\item  $w_I^1 = x_1 \oplus x_2 \oplus x_3.$
					
					\item $ w_I^2 = x_4 \oplus x_5 \oplus x_6 \oplus x_7 \oplus x_8.$
					
					\item  $N_{G_K^{2}}^{+}(9)=\{5,6,7,8\}, \\N_{G_K^{2}}^{+}(14)=\{4,5,6\},\\ N_{G_K^{2}}^{+}(13)=\{14\}$ and \\$N_{G_K^{2}}^{+}(11)=\{7,8\}$.\\ $w_{9}= x_{9} \oplus x_{5} \oplus x_6 \oplus x_7 \oplus x_8, \\w_{14}= x_{14} \oplus x_4 \oplus x_5 \oplus x_6 , \\w_{13}=x_{13} \oplus x_{14}$ and\\ $w_{11}=x_{11} \oplus x_7 \oplus x_8.$
					
					\item For $C_2 \in C^{S}$, an index code $\{ x_{15} \oplus x_{16}, x_{16} \oplus x_{17}, x_{17} \oplus x_{18} \}$ of length $3$ is obtained.
					
					\item For $C_3 \in C^{S}$, an index code $\{ x_{21} \oplus x_{22},x_{22} \oplus x_{23}, x_{23} \oplus x_{24} \}$ of length $3$ is obtained.
					
					\item For $C_4 \in C^{S}$, an index code $\{ x_{19} \oplus x_{20}, x_{20} \oplus x_{25}, x_{25} \oplus x_{26}\}$ of length $3$ is obtained.
					
					\item For each $j \in \{10,12\},\\$ $w_j=x_j.$
					
				\end{enumerate}

			 Number of transmitted symbols is $17$, which achieves the lower bound on the optimal broadcast rate. Hence this scheme is optimal.
			
			This IC structure obeys Condition \ref{cond: ncyc2} and we have shown that the index code constructed using Algorithm \ref{algo2} is optimal for this example. \label{exm: n_cyc_example5}
		\end{exmp}

	\section{Conclusion}
		In this work, we have studied IC structures, with interlocked outer cycles, which obey either Condition \ref{cond: ncyc1} or Condition \ref{cond: ncyc2} given in section \ref{main_results}. We have shown that index codes constructed using Algorithm \ref{algo2} are optimal for such cases. An interesting direction of further research is to find the maximum number of disjoint cycles in the non-inner vertex set.
	
	
	\section*{Acknowledgment}
	This work was supported partly by the Science and Engineering Research Board (SERB) of Department of Science and Technology (DST), Government of India, through J. C. Bose National Fellowship to B. Sundar Rajan.
	

\end{document}